\documentclass[twocolumn,preprintnumbers,amsmath,amssymb,aps,prb,longbibliography]{revtex4-1}
\usepackage{graphicx}
\usepackage{color}
\usepackage{dcolumn}
\usepackage{bm}
\usepackage{hyperref}
\begin{document}

\title{Characterizing Skyrmion Flow Phases
with Principal Component Analysis}  
\author{
 C. J. O. Reichhardt$^{1}$,	
	D. McDermott$^{2}$, 
 and C. Reichhardt$^{1}$
} 
\affiliation{
$^1$Theoretical Division,  
Los Alamos National Laboratory, Los Alamos, New Mexico 87545 USA\\ 
$^2$X-Theoretical Design Division, Los Alamos National Laboratory, Los Alamos, New Mexico 87545 USA\\ 
}

\date{\today}
\begin{abstract}
Principal component analysis (PCA) is a powerful method that can identify patterns in large, complex data sets by constructing low-dimensional order parameters from higher-dimensional feature vectors. There are increasing efforts to use space-and-time-dependent PCA to detect transitions in nonequilibrium systems that are difficult to characterize with equilibrium methods. Here, we demonstrate that feature vectors incorporating the position and velocity information of driven skyrmions moving through random disorder permit PCA to resolve different types of disordered skyrmion motion as a function of driving force and the ratio of the Magnus force to the dissipation. Since the Magnus force creates gyroscopic motion and a finite Hall angle, skyrmions can exhibit a greater range of flow phases than what is observed in overdamped driven systems with quenched disorder. We show that in addition to identifying previously known skyrmion flow phases, PCA detects several additional phases, including different types of channel flow, moving fluids, and partially ordered states. Guided by the PCA analysis, we further characterize the disordered flow phases to elucidate the different microscopic dynamics and show that the changes in the PCA-derived order parameters can be connected to features in bulk transport measures, including the transverse and longitudinal velocity-force curves, differential conductivity, topological defect density, and changes in the skyrmion Hall angle as a function of drive. We discuss how asymmetric feature vectors can be used to improve the resolution of the PCA analysis, and how this technique can be extended to find disordered phases in other nonequilibrium systems with time-dependent dynamics.
\end{abstract}
\maketitle
    
\section{Introduction}

Nonequilibrium systems can undergo
spatial and temporal changes as a function of changing external parameters,
and one of the biggest challenges in the field
is to understand how to characterize the resulting
different types of nonequilibrium phases and the transitions between them.
In equilibrium systems, different states can be characterized by spatial symmetries, and transitions between these states can be understood in terms of the breaking of these symmetries.
When structural disorder is present, however, even for equilibrium systems
it can be difficult to determine
what quantities to measure and how particular measures correlate with
different states or transitions.
In nonequilibrium systems, the problem becomes even more challenging
since the states often do not have long-range order and
can also change with time.

A powerful tool that can be used to identify patterns in large and complex data sets is
principal component analysis (PCA),
an unsupervised machine learning technique \cite{Abdi10,Shlens14}.
This method maps a high-dimensional set of feature vectors onto a much
lower dimensional set of order parameters that best explain the variance
present in the data.
PCA has been applied to biological systems \cite{McKinney06}
and used for pattern recognition \cite{Bishop06}.
It has also been used to 
identify phases and phase transitions
in condensed matter systems \cite{Wang16,Carrasquilla17,Wetzel17,Hu17}.
In three-dimensional systems of hard disks and of patchy colloidal particles,
a PCA-derived order parameter showed sharp changes across
equilibrium phase transitions \cite{Jadrich18}, and could detect
the different phases
without using standard measures such as structure factors.
PCA has also been used to find phase
transitions in spin systems \cite{Hu17}, in agreement
with standard phase transition measures but again without using the
standard measures.
These results suggest that PCA could be used to find
transitions among different states
in equilibrium systems under situations
where standard measures do not work or
when it is not known {\it a priori} what a good order parameter is.
In other work, PCA was shown to capture
signatures of different nonequilibrium states and the transitions between them
for several well-known models, including the random organization model
and a driven transition from
an isotropic to a nematic state in elliptical disks
\cite{Jadrich18a}.
This again suggests that a PCA analysis could be extended
to other nonequilibrium systems where different phases can arise.
McDermott {\it et al.} used a position-based PCA approach
to identify different types of flowing states for disks moving over random disorder \cite{McDermott20}.
A similar PCA analysis revealed the presence of
distinct motility-induced phase separation states in 
active matter \cite{McDermott23}.

A particularly well-studied example of a system that
undergoes transitions between different nonequilibrium states is
interacting particles driven over random quenched disorder.
The transitions can occur
between different types of pinned crystals, pinned glasses,
heterogeneous plastic flow phases, moving fluids, 
and moving crystals \cite{Fisher98,Reichhardt17}. 
For driven vortices in
type-II superconductors
\cite{Jensen88a,Bhattacharya93,Koshelev94,Higgins96,LeDoussal98,Balents98,Olson98a,Pardo98},
the depinning and sliding phases can be detected readily
in the transport signatures, which are proportional to
velocity versus force curves.
The different dynamic states produce
changes in the noise fluctuations and
in the structure, such as by altering
the number of particles with six neighbors and
the orientation of the topological defects.
Other systems that exhibit similar behavior include colloidal
particles moving over random substrates
\cite{Reichhardt02,Pertsinidis08,Tierno12a},
Wigner crystals in low-dimensional materials
\cite{Cha94,Reichhardt01,Madathil23},
active matter particles moving over disordered substrates
\cite{Morin17,Sandor17a},
driven pattern-forming systems \cite{Reichhardt03a,Zhao13},
driven emulsions \cite{LeBlay20},
flow in porous media \cite{Narayan94},
and frictional models \cite{Bylinskii16}.
Related behavior can occur for
sliding charge density waves \cite{Bhattacharya87,Brazovskii04},
interface depinning \cite{Leschhorn97},
dislocation depinning and motion  \cite{Moretti04, Vanossi12},
and earthquake models \cite{Fisher98}.
There is strong evidence that depinning transitions
can exhibit critical properties, associated with
the scaling of the velocity-force curves and
growing dynamical correlation lengths,
and that the system can show different scaling depending on whether the
depinning is plastic or elastic
\cite{Fisher98,Reichhardt17}.
For drives above depinning,
different types of flow states are possible,
including motion in quasi-1D channels where pinned and moving
particles coexist,
soliton-like flow, fluid-like flow, and polycrystalline flow.
At higher drives, the system can dynamically order into moving smectic or moving crystal phases.
In general, the three most clearly identifiable
phases are a pinned state at low drives, disordered or plastic flow at
intermediate drives,
and ordered flow at high drives.
There is some evidence that the plastic flow regime is not a
single type of flow but changes with increasing drive; however,
due to the strong spatial disorder present in the system during
plastic flow, traditional measures are unable to
distinguish between different disordered flow phases.

Recently, Reichhardt {\it et al.}~performed a particle-and-velocity-based (PVB)
principal component analysis (PCA) of
driven superconducting vortices moving over random disorder
\cite{Reichhardt25}.
This system was chosen because
it has been extensively studied in simulations and experiments and shown
to have transitions among
three well-known phases, pinned, plastic, and moving smectic,
as a function of quenched disorder strength
and external drive
\cite{Jensen88a,Bhattacharya93,Koshelev94,Higgins96,LeDoussal98,Balents98,Olson98a,Pardo98}.
It also has weak and strong pinning regimes.
For weak disorder, where the vortices
or particles retain their triangular lattice
structure,
the depinning is elastic and the particles keep their same neighbors
at depinning \cite{Fisher98,DiScala12},
while for strong disorder, the system breaks up at depinning.
Both pinning regimes produce distinctive features in the transport
curves and velocity distribution functions \cite{Reichhardt17}.
The three principal components of the PVB PCA were used to construct
three order parameters
that could identify the pinned, plastic, and moving smectic phases,
as well as several distinct disordered flow regimes.
In the plastic flow phase, PVB PCA was able to distinguish between what was
termed non-ergodic plastic flow, where the system is disordered
but certain particles or regions remain immobile or inaccessible,
and an ergodic flow regime, where the system is disordered but all
particles participate in the motion over time.
The ergodic phase could be further subdivided into
a phase where some particles are temporarily pinned and a
phase where all the particles are constantly moving but where
sufficiently strong velocity heterogeneities are present to maintain
the system in a disordered state.
In the non-ergodic plastic flow regime, the PVB PCA identified
a transition from quasi-one-dimensional (1D) plastic channels to
fully two-dimensional (2D) filamentary plastic flow.
The PVB PCA phase identification relied crucially on the fact that
information about both the positions and velocities of the particles was
included in the feature vector.
Once the different flow states had been identified with
PVB PCA, it was possible to compare the flow behavior with features in
the transport curves.

Another type of particle-like system that can pass through different
dynamical phases when driven over quenched disorder is
magnetic skyrmions, which are topological magnetic textures
that were discovered in chiral magnets in 2009
with neutron scattering measurements \cite{Muhlbauer09}.
The skyrmions were subsequently observed
using direct imaging techniques \cite{Yu10}.
Skyrmions form a triangular lattice and can be driven with applied currents
\cite{Jonietz10,Zang11,Schulz12,Yu12,Iwasaki13,Lin13a,Nagaosa13,Jiang17a},
interact with defects, and exhibit depinning thresholds
\cite{Schulz12,Iwasaki13,Fernandes18,Xiong19,Litzius20,Reichhardt22a,Gruber22,Xie24}.
Different types of sliding dynamic skyrmion phases have also been observed
\cite{Iwasaki13,Reichhardt15,Koshibae18,Reichhardt22a,Mallick24,Song24,Raab24}. 
Due to their small size, stability, and ability to be manipulated
with external drives, skyrmions are promising candidates for 
various applications \cite{Finocchio16,Fert17,Wang22,Lee23,Gomes25}.
As a result,
there is significant interest in understanding the dynamics of
skyrmions that are
interacting
with pinning sites or nanostructures \cite{EverschorSitte18,Reichhardt22a}.  

Although magnetic skyrmions share many
similarities with vortices in type-II superconductors,
their dynamics can be significantly different due to the presence of a
strong Magnus component
that can induce gyroscopic motion
\cite{Nagaosa13,EverschorSitte14,Reichhardt22a,Yang24}.
The resulting Magnus force can cause the skyrmions
to move at a finite Hall angle with respect to the 
driving force \cite{Nagaosa13,EverschorSitte14,Brearton21,Yang24},
in contrast to the zero Hall angle
overdamped dynamics typically observed in previously
studied driven particle-like systems with quenched disorder.
The Magnus force can lead to interesting effects, such as a side-jump behavior
for skyrmions moving over pinning sites.
The side-jump distance increases as the skyrmions
travel more slowly, 
resulting in a drive-dependent skyrmion Hall angle
\cite{Reichhardt15,Reichhardt16,Jiang17,Litzius17,Legrand17,Kim17,Reichhardt18a,Juge19,Zeissler20}.
The Magnus force can also produce
a swirling motion of the skyrmions around pinning sites or defects
\cite{Reichhardt22a},
leading to partial clustering effects in the presence of strong quenched
disorder
\cite{Koshibae18,Reichhardt19b,Reichhardt22a}. 
Simulations comparing the driven dynamics of
skyrmions to that of vortices showed that each
system exhibited a pinned state,
plastic motion, and a high drive
dynamically reordered phase \cite{Reichhardt15,Reichhardt16,Diaz17}.
The skyrmions dynamically reorder into
a triangular lattice, rather than
the moving smectic state found for the vortices.
This was attributed to
the Magnus force rotating the pinning-induced skyrmion velocity
fluctuations into the direction perpendicular to the drive \cite{Diaz17}, 
giving a more isotropic effective drive-induced temperature
compared to the strongly anisotropic velocity fluctuations along
the driving direction caused by pinning in 
overdamped systems  
\cite{LeDoussal98,Balents98,Olson98a,Pardo98}.
The presence of a Magnus force also modifies the
velocity noise spectra \cite{Diaz17,Sato19}.
When the Magnus-to-damping ratio is increased,
a larger portion of the pinning-induced fluctuations 
get rotated perpendicular to the drive,
increasing the extent of the fluid flow regime
\cite{Reichhardt15,Reichhardt19b,Reichhardt22a}
such that for drives at which
overdamped vortices would form a smectic state,
the skyrmions form a partially ordered fluid. 
The different phase transitions in skyrmions are associated with
signatures in the velocity-force curves, and there can be distinct
features not only in the velocity along the driving direction but also
in the velocity perpendicular to the driving
\cite{Reichhardt15,Reichhardt22a}.
The behavior of the skyrmion Hall angle is
also correlated with the different types of motion. The Hall
angle is low or close to zero
in the strongly plastic flow state, increases in the moving liquid phase,
and saturates to a level
close to the intrinsic skyrmion Hall angle in the driven crystal phase
\cite{Reichhardt15,Reichhardt22a}. 

In this work, we construct position-and-velocity-based feature vectors
and perform PVB PCA for driven skyrmions moving over random
disorder at varied driving force and different Magnus to damping force
ratios.
We use the first three principal components as order parameters and
show that we can construct a dynamical phase diagram based on the
PVB PCA results.
Some of the features in the principal components,
including peaks, dips, and zero crossings,
are correlated with features in the longitudinal and transverse
velocity-force curves, differential transport,
fraction of topological defects, and changes in the skyrmion Hall effect.
We found that by adjusting
the relative amount of position and velocity information in the feature
vector, we could improve the ability of PVB PCA to resolve different plastic
flow states relative to our original studies on overdamped
systems \cite{Reichhardt25}.
We show that PVB PCA successfully identifies the dynamic phases known
from previous work on driven skyrmions,
including a pinned phase, a plastic flow regime,
an extended moving fluid regime, and a moving crystal phase.
In addition, PVB PCA uncovers several new regions inside the plastic flow
state, 
including non-ergodic and ergodic plastic flow phases where the
flow channels change in structure and also undergo a tilting
due to the skyrmion Hall effect.
At higher drives, PVB PCA distinguishes two different states in the extended
fluid regime: 
a dissipation-dominated fluid, and a Magnus-dominated fluid.
Since the Magnus force generates local rotations that create
topological defects, the system cannot completely
recrystallize when the Magnus to damping force ratio is large,
and the drive needed to reach the reordered state increases
with increasing Magnus to damping force ratio.
In contrast, at lower Magnus to damping force ratios,
there is a competition between a tendency to form a smectic state
that is aligned with the driving force and a tendency to form a crystal that
is rotated slightly perpendicularly to the driving force,
leading to the appearance of a partially ordered state.

\section{METHODS}
\subsection{Simulation}

We model $N$ skyrmions in a 2D system of size $L \times L$ using
a modified Thiele equation approach
\cite{Lin13,Reichhardt15,Stidham20,Reichhardt22a,Reichhardt23,Brems25}.
The system has periodic boundary conditions in the $x$ and $y$
directions with $L=36\lambda$, where
$\lambda$ is the effective skyrmion diameter.
The skyrmion density is given by $\rho = N/L^2$.
We fix $N=1216$ and $\rho=1.0$ throughout this work.
The motion for skyrmion $i$ is numerically integrated using the following overdamped equation:
\begin{equation}
\alpha_d \mathbf{v}_{i} + \alpha_m \hat{\mathbf{z}}\times \mathbf{v}_{i} = \mathbf{F}^{ss}_{i} + \mathbf{F}^{sp}_i + \mathbf{F}^{D} .
\end{equation}
Here, $\mathbf{v}_{i} = d\mathbf{r}_{i}/dt$, where $\mathbf{r}_{i}$ is the position of skyrmion $i$.
The first term on the left represents the damping force,
which aligns the velocity with the direction of the net external forces.
The second term is the Magnus force,
which aligns the velocity perpendicular to the net external forces.
The repulsive skyrmion-skyrmion interaction force is given by
$\mathbf{F}_{i}^{ss} = \sum^{N}_{j\neq i}F_{0}K_{1}(r_{ij}/\lambda)\hat{\mathbf{r}}_{ij}$,
where $F_{0}$ is a force coefficient that can depend on material parameters, $r_{ij} = |\mathbf{r}_{i} - \mathbf{r}_{j}|$ is the distance between particles $i$ and $j$, and
$\hat{\mathbf{r}}_{ij}=(\mathbf{r}_i-\mathbf{r}_j)/r_{ij}$. The interaction is described by the modified Bessel function of the first kind, $K_1(r)$, which decays exponentially at large $R$, 
as derived from continuum-based models \cite{Lin13}.

The force from the quenched disorder, $\mathbf{F}^{sp}$,
is produced by $N_{\rm pin}$ randomly placed
non-overlapping pinning sites modeled as
attractive parabolic potential wells
with a maximum range of $R_p = 0.3\lambda$
and a maximum strength of $F_p$.
The pinning force is $\mathbf{F}_i^{sp} = \sum_{k=1}^{N_{\rm pin}} (F_p/R_p) \Theta(|\mathbf{r}_{ik}^{(p)}| - R_p) \hat{\mathbf{r}}_{ik}^{(p)}$,
where $\mathbf{r}_{ik}^{(p)} = \mathbf{r}_i - \mathbf{r}_k^{(p)}$
is the distance between vortex $i$ and pin $k$,
$\hat{\mathbf{r}}_{ik}^{(p)} = (\mathbf{r}_i - \mathbf{r}_k^{(p)})/|\mathbf{r}_{ik}^{(p)}|$, and $\Theta$ is the Heaviside step function.
Throughout this work we fix $F_p=1.0$. The pinning density
$\rho_p=N_{\rm pin}/L^2$ is fixed at $\rho_p=0.5$, the same value used
in Ref.~\cite{Reichhardt25}.
The external driving force is applied on
all the skyrmions in the $x$-direction, $\mathbf{F}^D = F_D \hat{\mathbf{x}}$.
We start from $F_D = 0.0$ and increase the drive in increments of $\Delta F_D = 0.005F_p$, spending $1 \times 10^5$ simulation time steps at each increment. The data is written out every 1000 simulation time steps, giving
$N_f = 100$ frames of data for every value of the drive.

To characterize the dynamic flow, we perform standard measures
including the average velocity
parallel,
$\langle V_{x}\rangle = \langle N^{-1} \sum_i^N v_i \cdot \hat{\mathbf{x}}\rangle$,
and perpendicular,
$\langle V_{y}\rangle = \langle N^{-1} \sum_i^N v_i \cdot \hat{\mathbf{y}}\rangle$, to the driving direction,
where the averaging is performed over each value of $F_D$.
The average fraction of sixfold coordinated particles is
$\langle p_6\rangle = \langle N^{-1} \sum_i^N \delta(z_i-6)\rangle$,
where $z_i$ is the coordination number of
skyrmion $i$ obtained from a Voronoi tessellation.
For a perfect triangular lattice, $p_6 = 1.0$.
Throughout this work, we fix $\alpha_d^2 + \alpha_m^2 = 1.0$
for varied values of $\alpha_d$ and $\alpha_m$.
This same normalization was used in previous work to allow for a
comparison of the net velocity-force curves
$\langle V \rangle$ versus $F_D$, where
$\langle V \rangle = \sqrt{\langle V_{x} \rangle^2 + \langle V_{x} \rangle^2}$.
For a system with no pinning, $\langle V \rangle = F_D$ for all drives
and the velocity-force curve is linear.
We also measure the differential
velocities $d\langle V_{x} \rangle/dF_D$ and $d\langle V_{y} \rangle/dF_D$
versus $F_D$.
For an overdamped system, features in the
velocity-force and differential transport curves have been
correlated with changes in the dynamical flow phases.
The intrinsic skyrmion Hall angle is
$\theta_{\text{intr}} = \tan^{-1}(\alpha_m/\alpha_d)$;
however, when disorder is present,
there is a finite driving threshold for motion,
the velocity-force curves become nonlinear,
and the measured skyrmion Hall angle $\theta_{sk}$ develops
a drive dependence such that
$\langle \theta_{sk}\rangle = \tan^{-1}(\langle V_{x} \rangle/\langle V_{y} \rangle)$.

\subsection{Principal Component Analysis}

For the principal component analysis, we use
an improvement of the position-and-velocity-based feature vectors developed in
our previous study \cite{Reichhardt25}.
PCA identifies the directions of maximum variance in the
space defined by the feature vectors \cite{Abdi10}.
In this way, a large amount of data can be reduced into a low dimensional
representation.
The maximum variance can be associated with the physical features
of the system and can show signatures
when the spatial pattern and the dynamics are changing in
character,
which could correspond to a shift from one
non-equilibrium phase to another.
We use the particle-based PCA approach
originally developed for off-lattice non-equilibrium systems
\cite{Jadrich18}, which has been used to identify a variety of
phases and transitions  \cite{Jadrich18,McDermott20,McDermott23}. 
The difference between the present work and
the PVB PCA analysis applied to driven superconducting vortex systems
\cite{Reichhardt25}
is a modification of the PVB PCA feature vector and
the introduction of a Magnus force $\alpha_m$.
In the limit of $\alpha_{m} = 0.0$, the behavior becomes equivalent
to that of a superconducting vortex system.

To construct our position-and-velocity-based feature vector, from
individual frames of the simulation we first randomly select $N_p$ probe
skyrmions and begin with the position component.
For each probe particle,
we compute the distance from the probe particle to all of the other
particles, $r_{ij} = |\mathbf{r}_{i}-\mathbf{r}_j|$,
and then sort these distances in ascending order.
We then place only the $n$ smallest distances
into an array in increasing order of distance:
\begin{equation} \vec{F}_i^{\rm raw} = [r_{i0}, r_{i1}, r_{i2}, ..., r_{ij}, ..., r_{in}] \label{eq:feature} \end{equation}
where $r_{i0} < r_{i1} < ... < r_{in}$.
The neighbor particles that appear in the
array form the neighbor set $\mathcal{N}_i$ for probe
particle $i$.
Importantly, from the neighbor set $\mathcal{N}_i$, we also construct a
reduced neighbor set $\mathcal{M}_i$ containing only the $m$ smallest
distances, with $m<n$. The second neighbor set will be used below.
This procedure is repeated for all $N_p$ probe particles.
From the resulting arrays,
we obtain the average distance to neighbor $j$ in frame $k$ of the
movie as $\langle r_{j}\rangle=N_p^{-1}\sum_i^{N_p} r_{ij}$.
We normalize the averaged distances according to
$\bar r_j=(\langle r_j\rangle - \mu_r)/\sigma_r$
where $\mu_r$ is the mean
and $\sigma_r$ is the standard deviation
of the $N_p$ values of $\langle r_j\rangle$.

To construct the velocity component of the feature vector, for each
probe particle we work only with the reduced neighbor set $\mathcal{M}_i$
identified above. We determine the net velocity $V_i=\sqrt{v_{xi}^2+v_{yi}^2}$
of each neighbor particle, and sort the velocities into an array in increasing
order:
\begin{equation} \vec{G}_i^{\rm raw} = [V_{i0}, V_{i1}, V_{i2}, ..., V_{ij}, ..., V_{im}] \label{eq:feature} \end{equation}
where $V_{i0} < V_{i1} < ... < V_{im}$.
We repeat this procedure for all $N_p$ probe particles and obtain the
average velocity of neighbor $j$ in frame $k$ of the movie as
$\langle V_j\rangle = N_p^{-1}\sum_i^{N_p} V_{ij}$.
The average velocities are normalized according to
$\bar V_j=(\langle V_j\rangle - \mu_V)/\sigma_V$ where $\mu_V$ is the mean
and $\sigma_V$ is the standard deviation of the $N_p$ values of
$\langle V_j\rangle$.

The position and velocity information is then assembled into an asymmetric
feature vector for each frame, with the position information first,
followed by the velocity information:
\begin{equation}
  \begin{split}
    \vec{f}_k = & [\bar{r}_{k0}, \bar{r}_{k1}, \bar{r}_{k2}, ..., \bar{r}_{kj}, ..., \bar{r}_{kn},\\
      & \bar{V}_{k0}, \bar{V}_{k1}, \bar{V}_{k2}, ..., \bar{V}_{kj}, ..., \bar{V}_{km}] \end{split} \label{eq:feature} \end{equation}
where $\bar{r}_{k0} < \bar{r}_{k1} < ... < \bar{r}_{kn}$
and $\bar{V}_{k0} < \bar{V}_{k1} < ... < \bar{V}_{km}$.
We construct a matrix $\mathbf{M}$ in which each row is the vector $\vec{f}_k$ obtained by processing each frame from the entire simulation data set. There are a total of $8 \times 10^4$ rows in matrix $\mathbf{M}$, since there are $N_f=100$ frames per current and there are 800 current values in the velocity sweep.
We find that for the skyrmion system, as for the superconducting vortex
system, the relatively
long-range particle-particle interactions smooth the density
fluctuations enough that it is not necessary to prewhiten the feature
vector, a procedure used in previous work for
particles with short-range interactions
to remove spurious geometric information caused by
strong particle clustering \cite{Jadrich18, McDermott20, McDermott23}.

We tested a variety of PVB feature vectors for the skyrmion system before
determining that an asymmetric feature vector based on distances and absolute
velocity gave the best resolution of the dynamic states. In our previous
study \cite{Reichhardt25}, the feature vector was symmetric and contained
$n$ entries for each of the distance, $x$ velocity, and $y$ velocity
components. The first issue with using a similar feature vector for the
skyrmion system is the nonzero Magnus force on the skyrmions. In the
overdamped superconducting vortex system, the particles always move
parallel to the $x$ direction driving force, so a separation of the velocity
into $x$ and $y$ components provided information on the motion parallel
and perpendicular to the drive. In the skyrmion system, the particles
move at a drive-dependent skyrmion Hall angle to the $x$ direction driving
force, meaning that as the drive increases, the $x$ and $y$ components of
the velocity contain blended amounts of
the motion parallel and perpendicular to the effective driving direction.
This blurred the reordering transition.
The blurring is much more noticeable in the skyrmion
system than in the superconducting vortex system, since the skyrmions have
a relatively sharp reordering transition.
To remedy this,
we first tested breaking the
velocities into the true parallel and perpendicular components at each
drive. This was achieved by determining the average direction of travel
in the frames belonging to a given current value, and then rotating our
velocity reference frame into the travel direction. This rather complicated
procedure gave some improvement but the reordering transition
was still blurred. As will be described in Sec. III B,
we later used the calculation
of the average direction and speed of travel at each
current for the purposes of image generation.

Still working with a symmetric feature vector, in which
both the distances and velocities contained $n$ entries,
we tested a variety of choices of $n$ and arrived at the
following insight:
Using larger $n$, which is equivalent to
running the feature vector out to larger distances, gives
improved resolution of the crystalline structure of the lattice, but
washes out information on the plastic distortions of the moving particles.
In contrast, using smaller $n$, which restricts the feature vector to
smaller distances, gives good resolution of the local plastic
distortions but blurs out information on the longer range crystalline
ordering of the lattice. Since there is no rule stating that the feature
vector must be symmetric, we settled on an asymmetric feature vector in
which the number of distance components $n$ is larger than the number of
velocity components $m$. This gives us good resolution of the crystalline
structure from the distance component, and good resolution of the plastic
flow behavior from the velocity component.
We also found that, with the asymmetric feature
vector, it is sufficient to use only the average velocity $V$, and 
breaking the velocity into parallel and perpendicular components is not
necessary. Finally, with our improved feature vector, we were able to
train the PVB PCA model on only a single value of $\alpha_m/\alpha_d$
and apply this pretrained
model to all other values of $\alpha_m/\alpha_d$.

In this work, we use $n=60$ distance neighbors,
$m=18$ velocity neighbors, and $N_p=100$ probe particles.
We use standard PCA techniques to compute the
orthogonal transformation matrix 
${\bf W}$ mapping our feature vectors of length $s=n+m$
to the principal components
based on data from the entire simulated velocity-force sweep for
a particular value of $\alpha_m/\alpha_d$,
 \begin{equation}
      {\vec p}_s = {\bf W}{\vec f}_s.
 \end{equation}
Here
${\vec p}_s$ are the principal components and
${\bf W} \equiv [{\vec q}_i, ..., {\vec q}_m]^T$, where the unit
vectors
${\vec q}_i$ define the directions of the principal components.
We define the PCA order parameters
  for a current $F_D$ from the average values of the principal component
  $\vec{p}_\alpha$ at that current according to:
    \begin{align}
  P_\alpha(F_D) &= \langle{\vec p}_\alpha(F_D)\rangle/\sqrt{\lambda_\alpha}\\
  &= N_p^{-1}\sum_{k \in F_D^i}{\vec w}^T_\alpha{\vec f}_k/\sqrt{\lambda_\alpha} .
  \end{align}
The normalization factor is the square root of $\lambda_\alpha$,
the eigenvalue associated with principal component $\alpha$. We find
that the first three principal components contain relevant
information, so that
$\alpha=\{1, 2, 3\}$.
Note that the eigenvectors
${\vec q}_i$ are only defined to within a sign
and remain eigenvectors if they are reflected across the origin.
As a result, the order parameters are
{\it also only defined to within a sign},
and although there is significance to the order parameter passing through
zero, since it means that the
effective dimensionality of the data has been
reduced by one,
there is no special significance to the fact that the order parameter
is positive or negative overall.
The PVB PCA model is pretrained on the $\alpha_m/\alpha_d=1.0$ system and
then applied to all systems with different values of $\alpha_m/\alpha_d$.
Pretraining is crucial for studying large values of $\alpha_m/\alpha_d$
because in this regime,
the dynamic reordering transition shifts to very high driving
forces that fall outside of our simulation window.

\section{Results}
\subsection{Transport Curves and Topological Defects}

\begin{figure}
\includegraphics[width=\columnwidth]{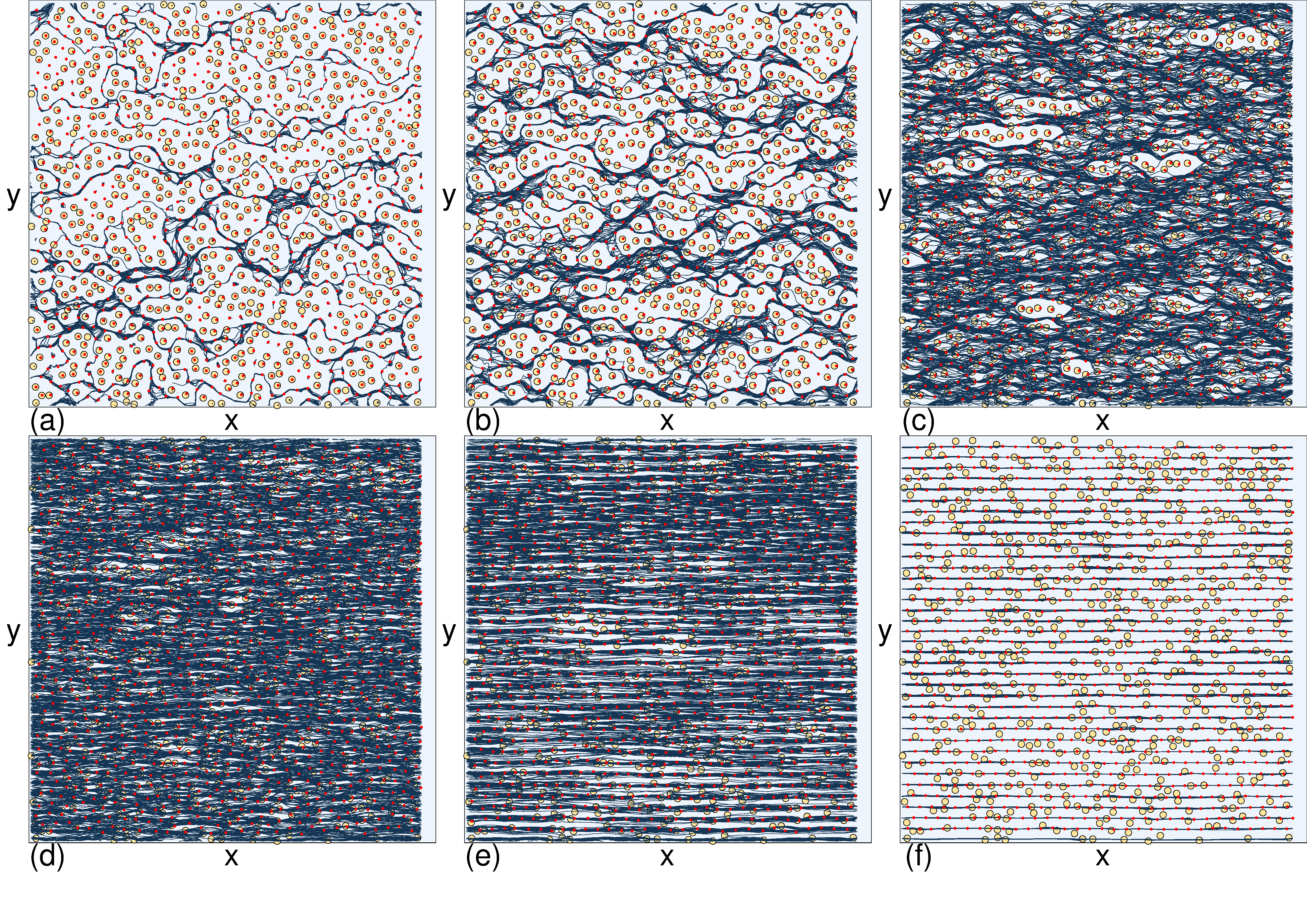}
\caption{Particle positions (dots), pinning sites (open circles), and trajectories (lines) for a system with $\alpha_m/\alpha_{d} = 0.0$ (the superconducting vortex limit) at different drives. The trajectories here and throughout
this work are imaged over a time
equal to that required for an individual particle moving through a pin-free
sample to travel a distance of $50\lambda$ when subjected to the driving
force $F_D$.
Phase I, the pinned state, is not shown.
(a) Phase II, the non-ergodic
isolated static channel flow phase, at
$F_D = 0.165$.
(b) Phase III, the lightly braided channel flow phase, at
$F_D = 0.31$.
(c) Phase IV, the heavily braided channel flow phase, at
$F_D = 0.48$.
(d) Phase V,the inhomogeneous ergodic plastic flow phase, at
$F_D = 0.61$.
(e) Phase VI, the emerging ordered flow phase, at
$F_D = 0.85$:
(f) DR, the dynamically reordered phase, at $F_D=2.0$. Here the vortices
dynamically reorder into a moving smectic state.
}
\label{fig:1}
\end{figure}

As a point of reference, we first illustrate the dynamic flow phases
identified with the assistance of PVB PCA in the superconducting vortex
system \cite{Reichhardt25}, in order to compare and contrast these states
with what we observe for the skyrmion system.
In Fig.~\ref{fig:1}, we plot the particle positions, trajectories,
and pinning site locations for a system with $\alpha_{m}/\alpha_{d} = 0.0$,
corresponding to the overdamped or superconducting vortex limit.
Phase I (not shown) is the pinned phase in which
there is no motion at a finite drive.
Figure~\ref{fig:1}(a) shows the motion at $F_{D} = 0.165$ in
phase II or the isolated channel flow state, 
a non-ergodic flow regime where a large number of particles both at
and between the pinning sites
do not take part in the motion, and the channel structure is static.
In Fig.~\ref{fig:1}(b) at $F_D=0.31$ in phase III or
the lightly braided channel flow phase,
the motion again occurs through channels, but now the
channels change gradually over time.
There are still particles both at the pinning sites and in the
interstitial regions that do not participate in the motion.
For $F_D=0.48$ in Fig.~\ref{fig:1}(c) in the heavily
braided channel flow phase IV,
the flow is more 2D and fluctuating in nature,
but there are still linear chains of pinned particles
that do not take part in the motion.
As a result, the dynamics remains non-ergodic, and the
velocity distribution function is bimodal with
finite weight at $\langle V_{x} \rangle = 0.0$.
Figure~\ref{fig:1}(d) shows $F_D=0.61$ in phase V or the inhomogeneous ergodic
plastic flow phase.
All the particles take part in the motion over time;
however, at any given moment there are still pinned particles present,
so the velocity distribution function remains bimodal.
The system can be regarded as a moving fluid.
In Fig.~\ref{fig:1}(e) at $F_{D}/F_{p} = 0.85$ we show phase VI,
the emerging ordered flow phase, where all the particles
are moving and local patches of topological order are present.
At $F_D/F_p=2.0$ in Fig.~\ref{fig:1}(f), the system has reached
a dynamically reordered (DR) state, which for the
superconducting vortex system is
a moving smectic. The particles move in well-defined 1D paths along
the driving direction,
and there are only a small number of topological defects
present in the form of dislocations that have their Burgers vectors
aligned parallel to the drive.
For higher drives, the smectic regime
persists since the particles remain locked in the flowing 1D channels.
As has been shown previously,
the sixfold order parameter $\langle p_{6}\rangle$ most clearly captures the
emergence of the moving smectic phase. The boundaries between
phases II, III, IV, V, and VI are difficult to distinguish from
transport and noise measures; however, we showed previously that
PVB PCA
produces three principal component
order parameters 
that readily identify the distinct plastic flow states \cite{Reichhardt25}. 

\begin{figure}
\includegraphics[width=\columnwidth]{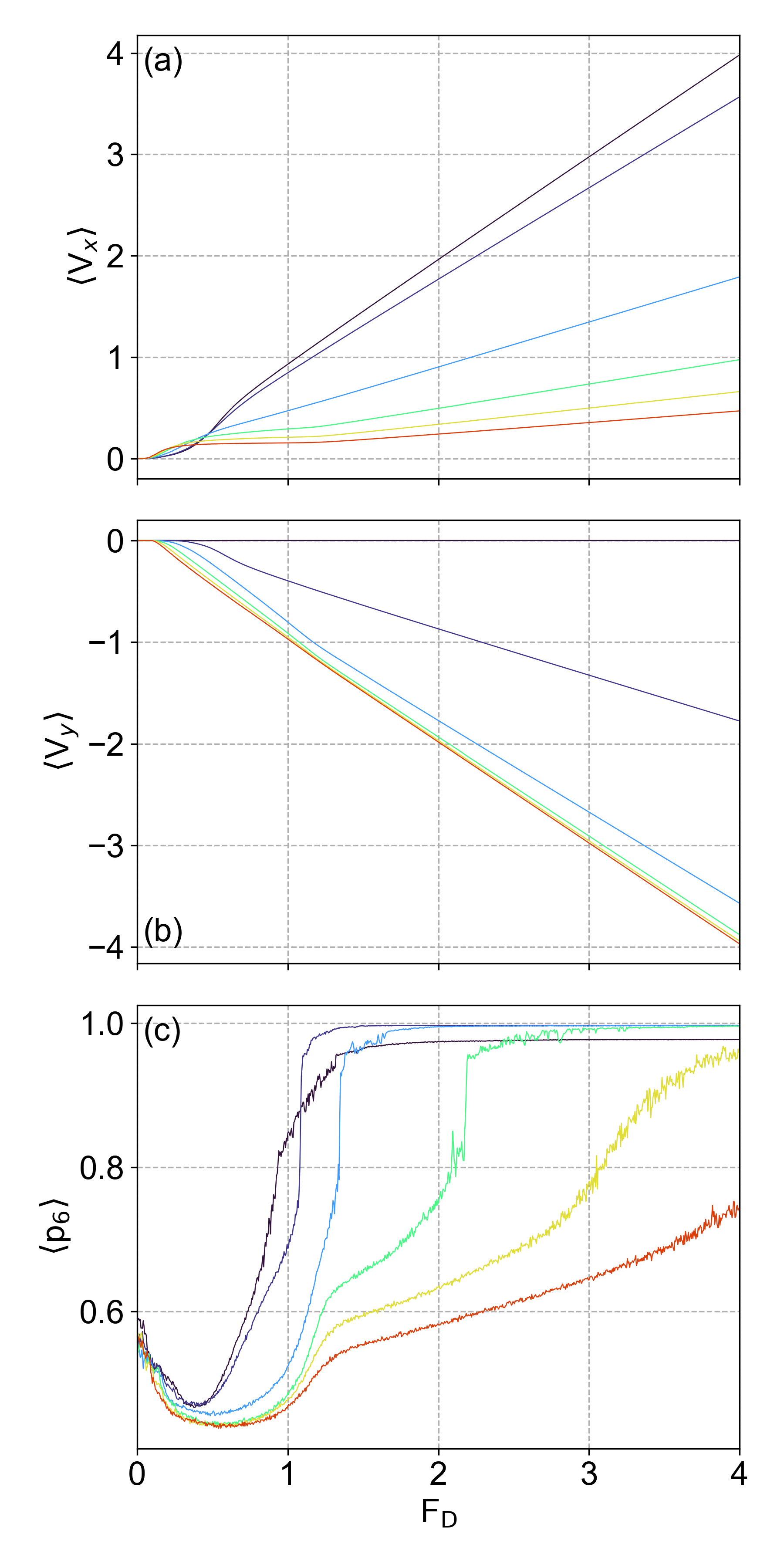}
\caption{(a) The longitudinal velocity $\langle V_x\rangle$ vs $F_D$
for systems with
$\alpha_m/\alpha_d = 0.0$ (black), 0.5 (purple),
2.0 (blue), 4.0 (green), 6.0 (yellow), and 8.5 (red).
(b) The corresponding transverse velocity $\langle V_y\rangle$ vs $F_D$.
(c) The corresponding
fraction of particles with six neighbors $\langle p_6\rangle$ vs $F_D$.}
\label{fig:2}
\end{figure}

\begin{figure}
\includegraphics[width=\columnwidth]{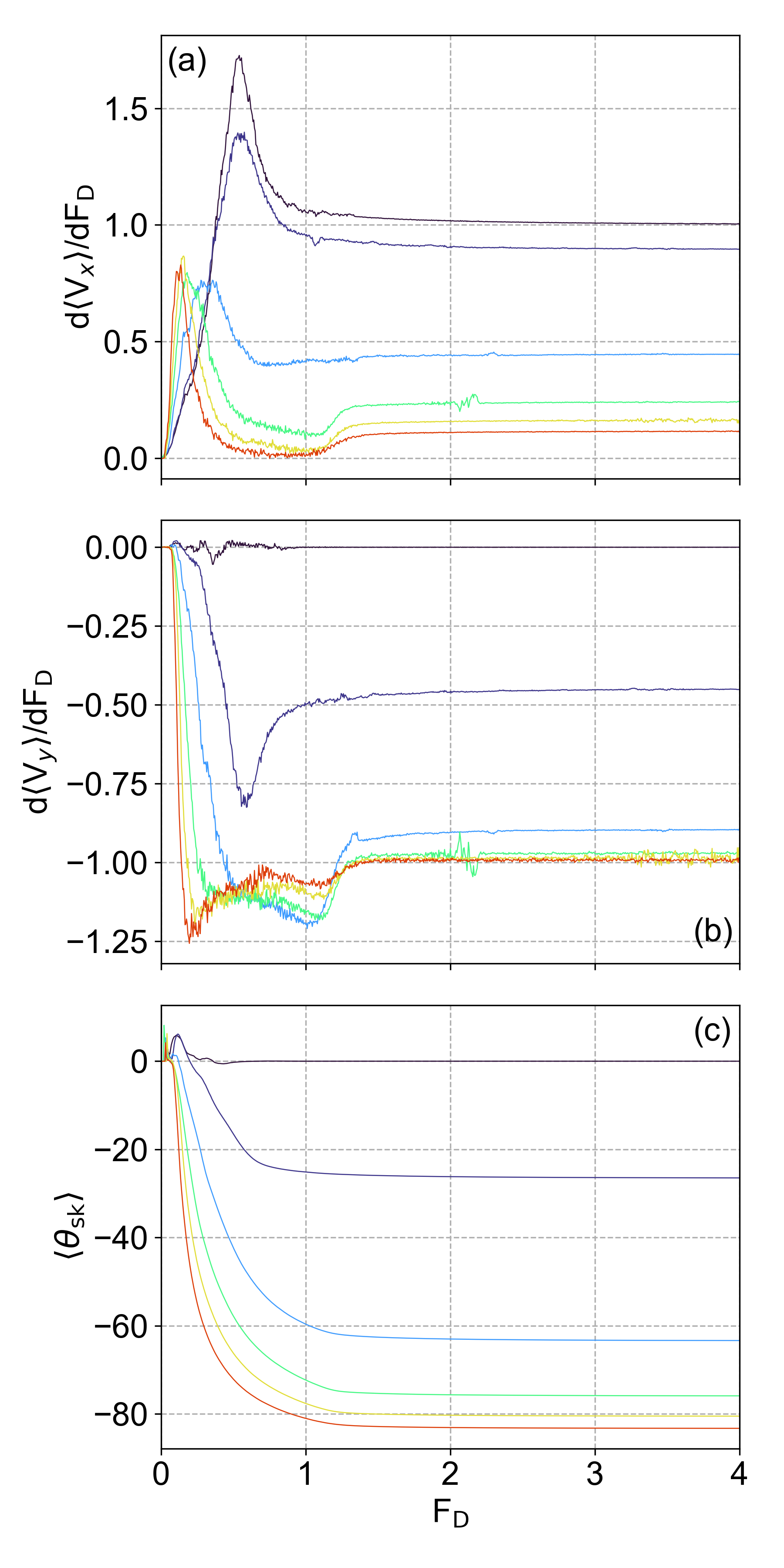}
\caption{(a) $d\langle V_x\rangle/dF_D$ vs $F_D$ for the samples from
Fig.~\ref{fig:2} with
$\alpha_m/\alpha_d = 0.0$ (black), 0.5 (purple),
2.0 (blue), 4.0 (green), 6.0 (yellow), and 8.5 (red).
(b) The corresponding $d\langle V_y\rangle/dF_D$ vs $F_D$.
(c) The corresponding average skyrmion Hall angle
$\langle \theta_{sk}\rangle$ vs $F_D$.
}
\label{fig:3}
\end{figure}

We next enter the
skyrmion regime by considering systems with nonzero
$\alpha_m/\alpha_d$.
In Figs.~\ref{fig:2}(a,b,c), we plot $\langle V_x\rangle$,
$\langle V_y\rangle$, and $\langle p_6\rangle$, respectively,
versus $F_D$ at $\alpha_m/\alpha_d = 0.0$, 0.5, 2.0, 4.0, 6.0, 
and 8.5.
Figures~\ref{fig:3}(a,b,c) show the corresponding
$d\langle V_x\rangle/dF_D$, $d\langle V_y\rangle/dF_D$,
and $\langle \theta_{sk}\rangle$ versus $F_D$ curves.
For $\alpha_m/\alpha_d = 0.5$, both $V_x$ and $V_y$ increase
with increasing $F_D$, so
there can be a finite Hall angle, as shown in Fig.~\ref{fig:3}(c).
The threshold for motion is higher in the $y$ direction than in the
$x$ direction, which is due to the fact that
just above depinning, the slowly moving skyrmions
initially remain confined by the filamentary 1D channels, giving
a zero Hall angle.
Previous simulation and experimental studies
also showed that above the depinning transition,
there is a range of drives for which the Hall angle is zero.
This is more significant at finite temperatures in the creep regime,
where the particles are able to hop between pinning sites;
however, we remain in the zero temperature limit in the
present work.
The $\langle p_6\rangle$ versus $F_D$ curve at
$\alpha_m/\alpha_d = 0.5$ follows the $\alpha_m/\alpha_d$ curve relatively
closely but shows two significant differences: for
$\alpha_m/\alpha_d=0.5$,
there is a much sharper jump into the ordered phase,
and $\langle p_6\rangle$ reaches a value that is much closer to
$\langle p_6\rangle=1.0$ in the dynamically reordered (DR) state at
higher drives.
This indicates that
the system is more topologically ordered in the DR phase compared to the
$\alpha_m/\alpha_d = 0.0$ system, since the skyrmion system with
$\alpha_m/\alpha_d=0.5$ forms a reordered moving crystal
rather than the reordered moving smectic that appears for the
$\alpha_m/\alpha_d=0.0$ vortex system.

At $\alpha_m/\alpha_d=0.5$,
both the $d\langle V_x\rangle/dF_D$ versus $F_D$ curve in
Fig.~\ref{fig:3}(a) and the
$d\langle V_y\rangle/dF_D$ versus $F_D$ curve in
Fig.~\ref{fig:3}(b) pass through a peak or minimum, respectively,
before reaching a saturation value at large drives,
while in Fig.~\ref{fig:3}(c),
the skyrmion Hall angle
$\langle \theta_{sk}\rangle$
starts off at zero just above depinning and then linearly increases
in magnitude with increasing $F_D$,
showing a saturation that begins near $F_D = 0.8$
in which $\langle \theta_{sk}\rangle$
approaches the expected intrinsic value of $\theta_{\rm intr}=-26.56^\circ$.
The drive dependence of the skyrmion Hall angle in the plastic flow
phase has been studied in previous experiments and simulations,
where a similar linear increase of the Hall angle magnitude to the intrinsic
Hall angle was observed as a function of drive
over a similar range of values of $\alpha_m/\alpha_d.$

\begin{figure}
\includegraphics[width=\columnwidth]{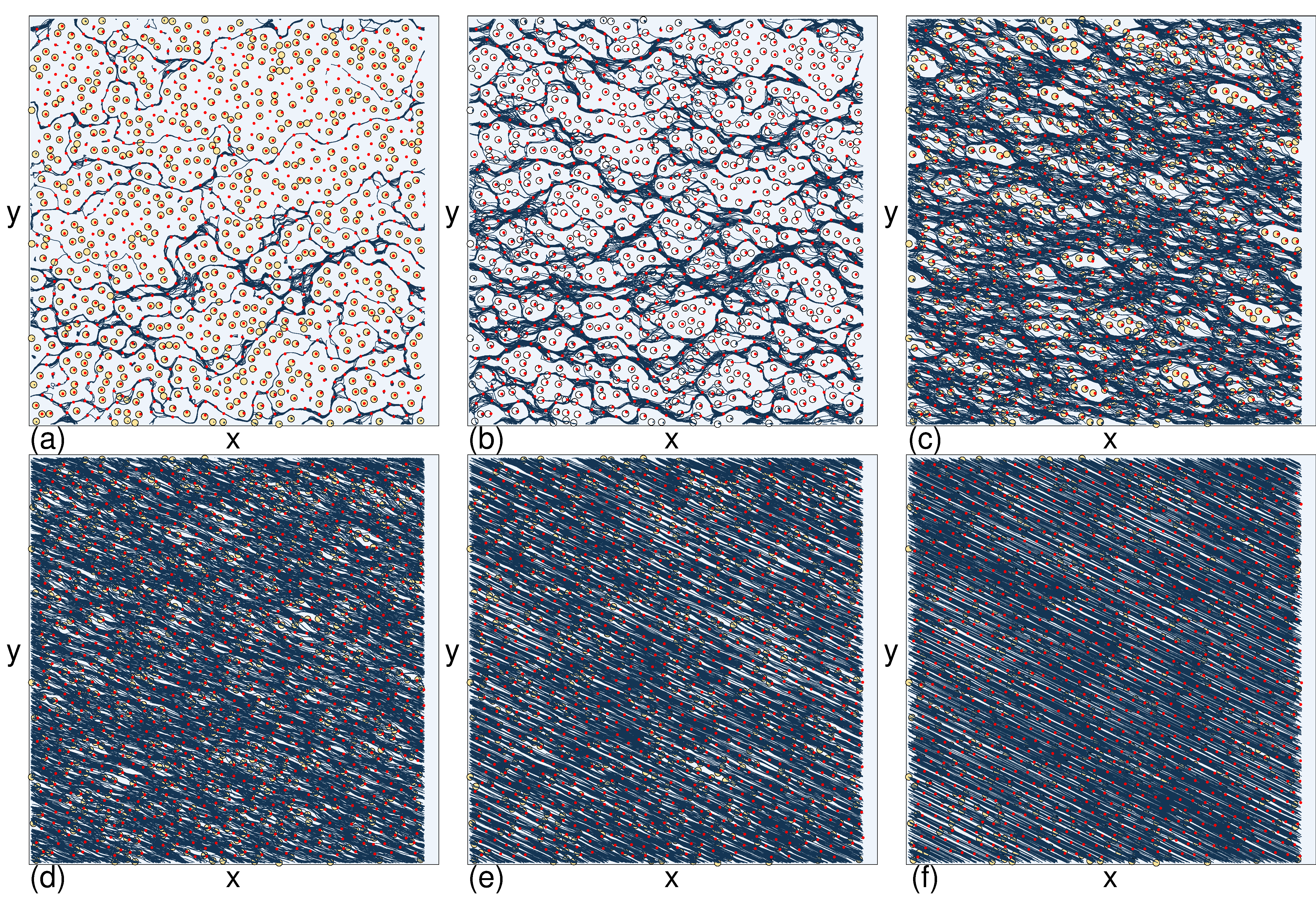}
\caption{Particle positions (dots), pinning sites (open circles), and
trajectories for a system with $\alpha_m/\alpha_d = 0.5$.
(a) Phase II, isolated channel flow, at $F_D = 0.15$. Here the
Hall angle is zero. (b) Phase III, lightly braided channel flow,
at $F_D = 0.31$. There is now some evidence of tilt in the
trajectories due to the emergence of a finite Hall angle.
(c) Phase IV, heavily braided channel flow,
at $F_D = 0.48$. The Hall angle has increased, and there are still
some pinned particles present.
(d) Phase V, inhomogeneous ergodic plastic flow,
at $F_D = 0.63$. The system is in a moving fluid state,
but some particles can be temporarily pinned.
(e) Phase VI, emerging ordered flow,
at $F_D = 0.97$, where all particles are flowing at all times
but with varying speeds.
(f) The DR phase at $F_D = 2.0$, where the system forms a triangular lattice
moving at a finite Hall angle.
 }
\label{fig:4}
\end{figure}

In Fig.~\ref{fig:4}(a), we show the skyrmion positions, trajectories,
and pinning site locations in a system with
$\alpha_m/\alpha_d = 0.5$ at $F_D = 0.15$
in phase II, where the motion is in isolated static channels.
A large number of the particles are pinned at the pinning sites and in
the interstitial regions between the pinning sites.
At this drive,
$\langle \theta_{sk}\rangle = 0.0^\circ$, and there is no indication of
a nonzero Hall angle from the trajectories; all of the particles are
moving in the direction of the driving force.
Figure~\ref{fig:4}(b) shows phase III flow at $F_D=0.31$.
The Hall angle is now finite but small, much smaller
than the intrinsic value of $\theta_{\rm intr}=-26^\circ$.
In phase IV, Fig.~\ref{fig:4}(c) shows that at $F_D=0.48$,
there are still some permanently pinned particles,
but now the linear chains of pinned particles are tilted with respect to
the driving or $x$ direction as a result of the increasing
Hall angle.
At $F_D=0.63$ in phase V, shown in Fig.~\ref{fig:4}(d), the system is
topologically disordered and undergoes ergodic liquid like flow.
In Fig.~\ref{fig:4}(e) at $F_D=0.97$ we find phase VI flow,
where there is more channeling of the particle motion.
The flow clearly has a finite Hall angle.
Finally in Fig.~\ref{fig:4}(f) we show the DR phase at $F_D=2.0$,
where the particles have reordered into a triangular lattice.
In this case, we do not observe
the well-defined 1D channels of flow that appear
for $\alpha_m/\alpha_d = 0.0$.
This is because the skyrmion system forms a triangular floating solid rather
than a moving smectic state.
The smectic phase is still strongly coupled to the pinning,
while the moving crystal is far less affected by the pinning. In addition,
the linear change of the Hall angle with increasing drive tends to
destabilize the formation of channels since the channels keep ceasing
to be aligned with the direction of skyrmion motion as the drive becomes
larger.

For $\alpha_{m}/\alpha_{d} = 2.0$, in Fig.~\ref{fig:2}(b)
the $\langle V_y\rangle$ versus $F_D$ curve
becomes much more prominent,
and there is an upward shift
of the drive at which the system dynamically reorders,
indicated by the point where $\langle p_{6}\rangle$ approaches 1.0
in Fig.~\ref{fig:2}(c).
A shoulder feature begins to emerge in $\langle p_6\rangle$ above
$F_{D} > 1.0$, and there is a sharp jump
of $\langle p_6\rangle$ when the system reaches
the ordered phase.
There are also changes
in the $d\langle V_x\rangle/dF_{D}$ versus $F_D$ curve
shown in Fig. 3(a), including a clear shift in the peak to lower drives
compared to the $\alpha_m/\alpha_d=0.0$ system, as well as the first signs
of a shoulder above $F_D>1.0$.
The $d\langle V_y\rangle/dF_{D}$ versus $F_D$
curve in Fig. 3(b) now has a double dip feature, with a deeper dip
following a shallow dip,
while $\langle \theta_{sk}\rangle$ versus $F_{D}$ shows a saturation
to the intrinsic value of $\theta_{\rm intr}=-63.43^\circ$ at higher drives.
In general, for intrinsic Hall angles smaller in magnitude than
$|\theta_{\rm intr}|=45^\circ$,
there is a linear increase of
$\langle \theta_{sk}\rangle$ with increasing $F_{D}$,
while for higher intrinsic Hall angle magnitudes, the linearity is
lost and the curve becomes more quadratic at low drives.

\begin{figure}
\includegraphics[width=\columnwidth]{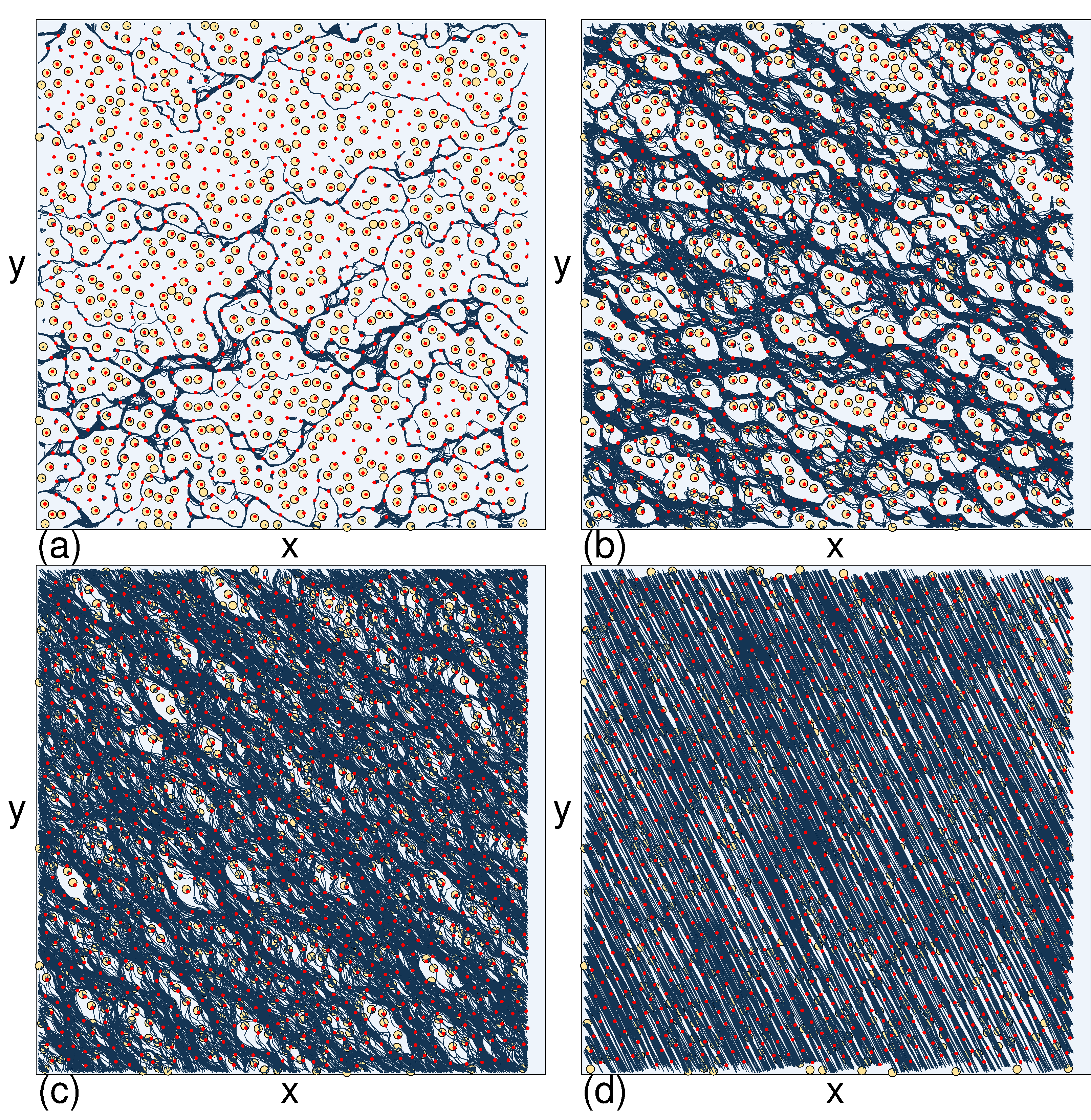}
\caption{Particle positions (dots), pinning site locations (open circles),
and trajectories (lines) for a system with $\alpha_m/\alpha_d = 2.0$.
(a) Phase II at $F_D=0.13$.
(b) Phase III at $F_D=0.36$.
(c) Phase IV at $F_D=0.6$.
(d) DR or moving crystal phase at $F_D=1.11$.  
}
\label{fig:5}
\end{figure}

In Fig.~\ref{fig:5}(a), we show the particle positions and trajectories
along with the pin locations for the $\alpha_m/\alpha_d = 2.0$
system in phase II at $F_D=0.13$.
This state resembles the phase II flow found at
lower values of $\alpha_m/\alpha_d$.
At $F_D=0.36$ in Fig.~\ref{fig:5}(b), the system is in phase III,
where the trajectories are more tilted due to the increasing Hall angle.
Figure~\ref{fig:5}(c) shows phase IV at $F_D=0.6$
with even stronger tilting of the trajectories,
while a diminished number of particles remain pinned.
Finally in Fig.~\ref{fig:5}(d),
at $F_D=1.11$ the system is in
the DR or moving crystal phase,
where there are no well-defined 1D flow channels
but the particles have formed
a triangular lattice.

The $\alpha_m/\alpha_d = 4.0$ curves show a strong
deviation from the $\alpha_m/\alpha_d = 0.0$ curves
in Figs.~\ref{fig:2} and \ref{fig:3}.
For $\alpha_m/\alpha_d=4.0$,
the $\langle p_6\rangle$ versus $F_D$ curve
shows a plateau-like region followed by
a sharp transition to the dynamically ordered phase above $F_D = 2.0$.
The differential transport curves
$d\langle V_x\rangle/dF_D$ and $d\langle V_y\rangle/dF_D$ versus
$F_D$ have a two-step feature,
and there is a small cusp near $F_D = 2.0$ that is associated with
the reordering transition.
The value of $\langle \theta_{sk}\rangle$ saturates near $F_D = 1.5$.
For $\alpha_m/\alpha_d = 6.0$, a similar set of features appear in the
transport curves,
but now the plateau in $\langle p_6\rangle$ extends out to higher drives.
The plateau region occurs for drives above the
reordering transition of the $\alpha_m/\alpha_d = 0.0$ system,
and on the plateau, $\langle p_6\rangle$ takes values
between $0.55$ and $0.9$,
indicating that the system is partially ordered but has not yet formed
a moving crystal. The
width of the plateau becomes larger as $\alpha_m/\alpha_d$ increases,
suggesting that it is produced by the Magnus force.

\begin{figure}
\includegraphics[width=\columnwidth]{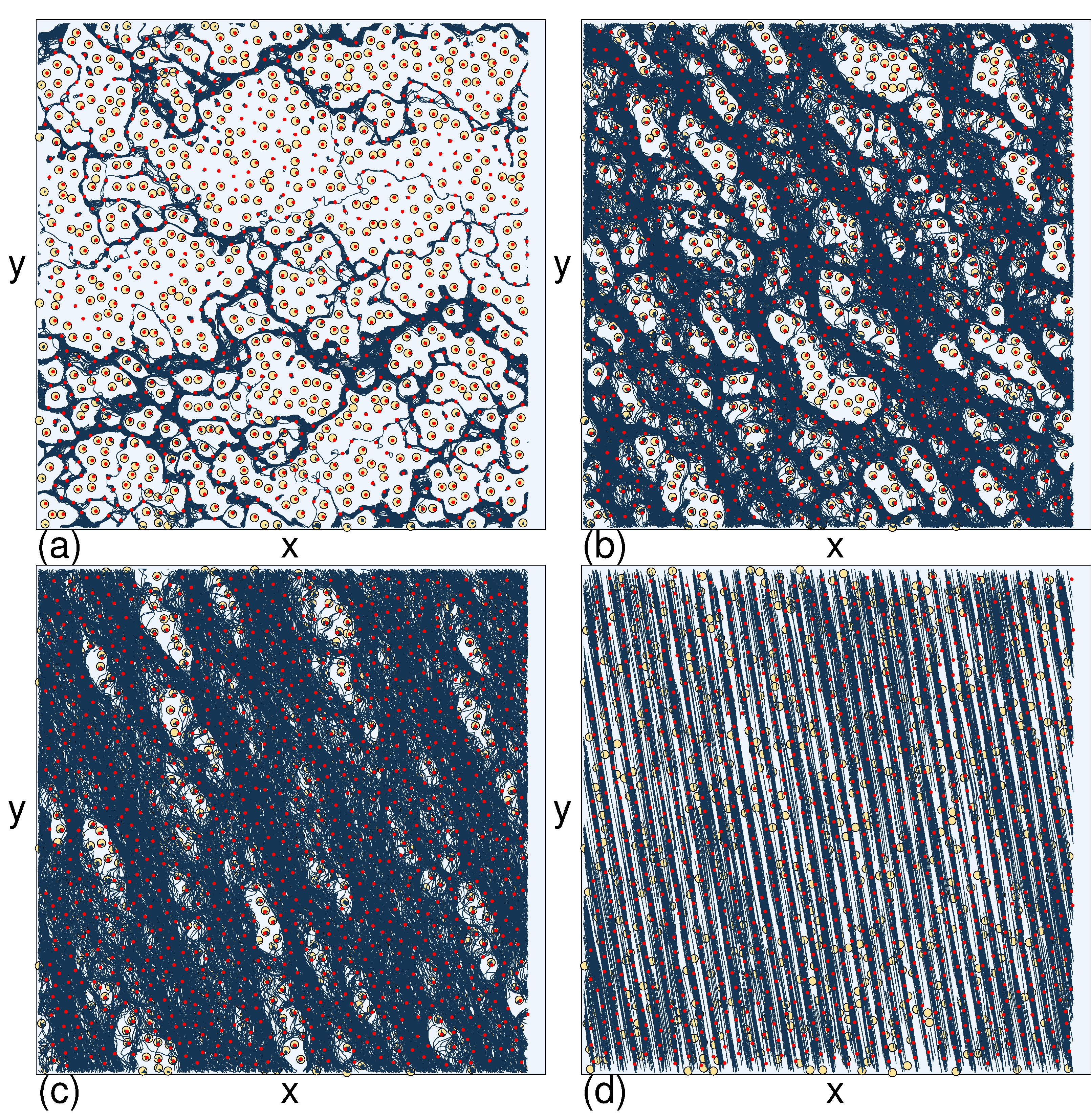}
\caption{Particle positions (dots), pinning site locations (open
circles), and trajectories (lines) for
a system with $\alpha_m/\alpha_d = 6.0$.
(a) Phase II at $F_D=0.12$. (b) Phase III at $F_D=0.34$.
(c) Phase IV at $F_D=0.64$. (d) DR or moving crystal phase at $F_D=4.0$.
}
\label{fig:6}
\end{figure}

In Fig.~\ref{fig:6}(a), we show the particle positions and trajectories
for the $\alpha_m/\alpha_d = 6.0$ system at $F_D=0.12$
in phase II,
where the Hall angle is much smaller than the intrinsic Hall angle.
In this case, the channels have become less 1D-like due to the fact that
the increased Magnus force creates more gyroscopic-like orbits.
There are, however, still well-defined channels
separated by large regions where motion does not occur.
Figure~\ref{fig:6}(b) illustrates the same system
at $F_D=0.34$ in phase III, where the channels are now more
strongly tilted and are also fatter than the phase III flow channels that
form for lower values of $\alpha_m/\alpha_d$.
In Fig.~\ref{fig:6}(c) at $F_D=0.64$ in phase IV,
there is stronger tilting of the channels,
while at $F_D=4.0$ in Fig.~\ref{fig:6}(d), the system is in the DR
moving crystal state.
The $\alpha_m/\alpha_d = 8.75$ system does not reach a dynamically ordered
state over the range of drives we consider, and
in Fig.~\ref{fig:2}(c) we find that
$\langle p_6\rangle = 0.73$ at $F_D = 4.0$
for this ratio of Magnus to damping force.
The trajectories at $\alpha_m/\alpha_d = 8.75$ in the plastic flow
states are
similar to those found for $\alpha_m/\alpha_d = 6.0$.

\begin{figure}
\includegraphics[width=\columnwidth]{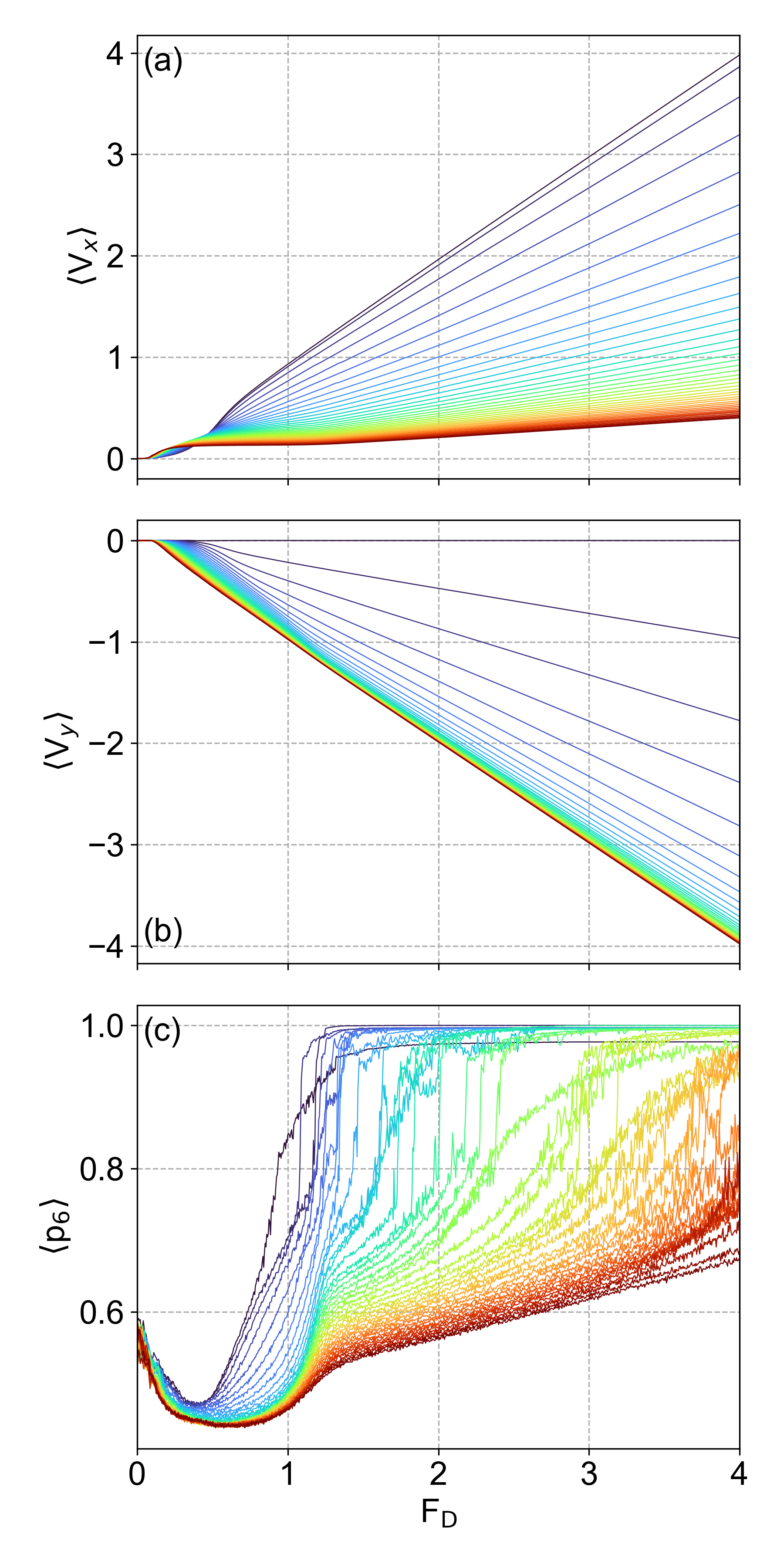}
\caption{(a) $\langle V_x\rangle$ vs $F_D$ at varied $\alpha_m/\alpha_d$
from $\alpha_m/\alpha_d=0.0$ (black) to $\alpha_m/\alpha_d=10.0$ (dark red)
in intervals of 0.25.
(b) The corresponding $\langle V_y\rangle$ vs $F_D$.
(c) The corresponding $\langle p_6\rangle$ vs $F_D$.
}
\label{fig:7}
\end{figure}

In Fig.~\ref{fig:7}, we plot $\langle V_x\rangle$, $\langle V_y\rangle$,
and $\langle p_6\rangle$ versus $F_D$ for varied
$\alpha_m/\alpha_d$ from 0.0 to 10.0 in increments of 0.25.
Here, it can be seen that there are
multiple crossings of the $V_x$ curves as $\alpha_m/\alpha_d$ increases,
and the drive at which the reordering transition occurs shifts
to higher values. The jump in $\langle p_6\rangle$ accompanying the reordering
transition is somewhat stochastic and can shift slightly from one disorder
realization to another, but the trend of increasing drive for the reordering
transition with increasing $\alpha_m/\alpha_d$ remains robust.
For the largest values of $\alpha_m/\alpha_d$,
there is a rapid increase
in $\langle p_6\rangle$ just above $F_D = 1.0$,
followed by a plateau region,
and finally by a jump at higher drives
corresponding to the point at which
the system enters the moving crystal DR phase.

\begin{figure}
\includegraphics[width=\columnwidth]{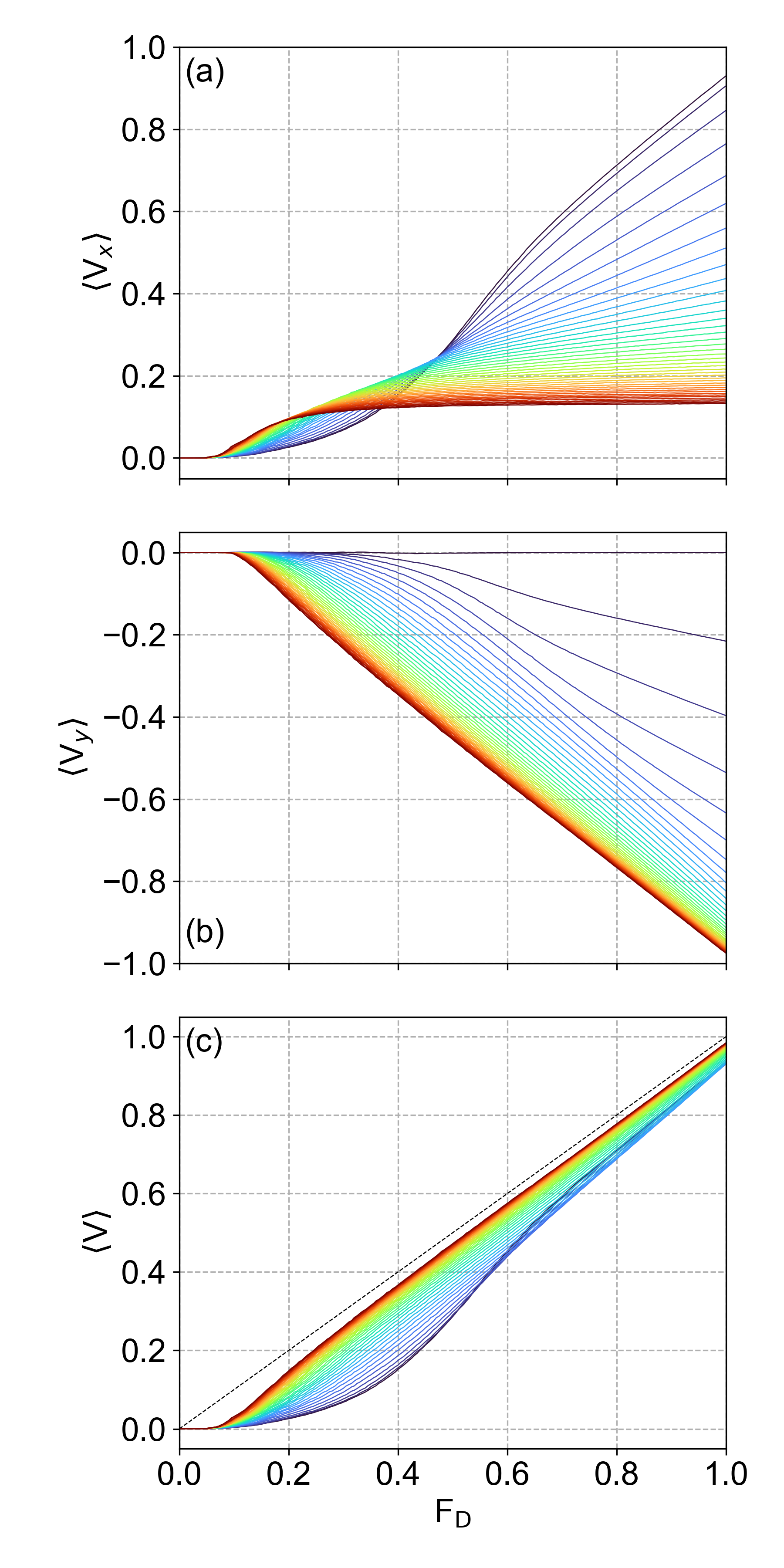}
\caption{(a) A blow-up of $\langle V_x\rangle$ vs $F_D$ for the system
from Fig.~\ref{fig:7},
showing the crossing of the velocity-force curves with
increasing $\alpha_m/\alpha_d$.
Here $\alpha_m/\alpha_d$ ranges from 0 (black) to 10.0 (dark red) in
intervals of 0.25.
(b) The corresponding $\langle V_y\rangle$ vs $F_D$ curves.
(c) The net velocity $V = \sqrt{V_x^2 + V_y^2}$ vs $F_D$,
where the dashed line indicates the response expected
in the pin-free limit.
There is a crossing of the curves for low $\alpha_m/\alpha_d$.
}
\label{fig:8}
\end{figure}

In Fig.~\ref{fig:8}(a,b), we show a blow-up of
the $\langle V_x\rangle$ and $\langle V_y\rangle$
versus $F_D$ curves from
Fig.~\ref{fig:7} up to a maximum value of $F_D = 1.0$,
which more clearly shows how the velocities evolve in the plastic
flow regime.
At lower $\alpha_m/\alpha_d$, $\langle V_x\rangle$
is low just above depinning and increases at higher drives.
In contrast, at higher $\alpha_m/\alpha_d$,
$\langle V_x\rangle$ increases rapidly above depinning
but then saturates with increasing drive,
which leads to a crossing of the
velocity-force curves between the low and high values of $\alpha_m/\alpha_d$.
The net velocity curves themselves cross above $F_D=0.5$ for lower
values of $\alpha_m/\alpha_d$, as shown in the plot of
$\langle V\rangle = \sqrt{\langle V_x\rangle^2+\langle V_y\rangle^2}$
versus $F_D$ in Fig.~\ref{fig:8}(c).
If increasing the relative importance of the Magnus
term simply weakened the effectiveness of the pinning, the
velocity-force curves would gradually approach the pin-free limit,
indicated by a dashed line in Fig.~\ref{fig:8}(c). This is true at
low $F_D$, but in the region where the velocity-force curves cross,
the pinning is effectively becoming stronger with increasing
$\alpha_m/\alpha_d$, not weaker. The crossing occurs because
the slope of $\langle V_x\rangle$ above $F_D=0.5$ rapidly decreases
with increasing $\alpha_m/\alpha_d$ at low $\alpha_m/\alpha_d$, but the
magnitude of the slope of $\langle V_y\rangle$ does not increase
rapidly enough to offset this decrease. As a result, the net velocity
above $F_D=0.5$ initially decreases as $\alpha_m/\alpha_d$ increases.
Eventually, the slope of $\langle V_x\rangle$ begins to saturate toward
a value of zero while the magnitude of the slope of $\langle V_y\rangle$
continues to increase with increasing $\alpha_m/\alpha_d$, and the net
velocity begins to increase with increasing $\alpha_m/\alpha_d$.

\begin{figure}
\includegraphics[width=\columnwidth]{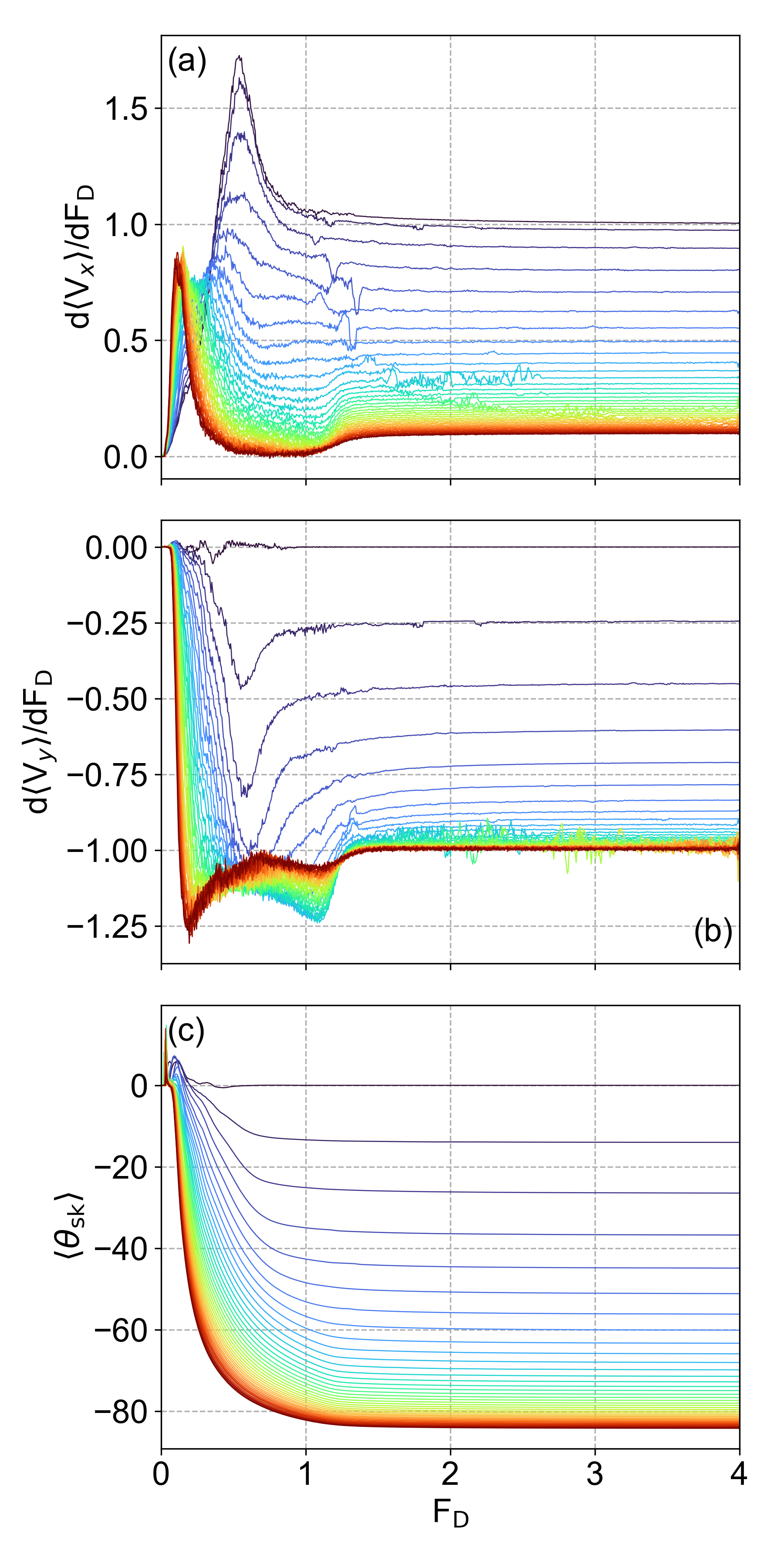}
\caption{(a) $d\langle V_x\rangle/dF_{D}$ vs $F_D$
for varied  $\alpha_m/\alpha_d$ over the range
$\alpha_m/\alpha_d=0.0$ (black) to $\alpha_m/\alpha_d=10.0$ (dark red)
in intervals of 0.25.
(b)  The corresponding $d\langle V_y\rangle/dF_{D}$ vs $F_{D}$.
(c) The corresponding $\langle \theta_{sk}\rangle$ vs $F_D$.}
\label{fig:9}
\end{figure}

In Fig.~\ref{fig:9}(a,b,c), we plot
$d\langle V_x\rangle/dF_D$, $d\langle V_y\rangle/dF_D$,
and $\langle \theta_{sk}\rangle$, respectively,
versus $F_D$ for varied $\alpha_m/\alpha_d$.
For lower $\alpha_d/\alpha_m$, $d\langle V_x\rangle/dF_D$
has a single strong peak.
As $\alpha_d/\alpha_m$ increases, this first peak weakens and a second,
smaller peak begins to emerge near $F_D = 1.2$. The first peak shifts to
lower $F_D$ and the second weak peak shifts to higher $F_D$ as
$\alpha_m/\alpha_d$ increases.
At high values of $\alpha_m/\alpha_d$, there is
a wide region where $d\langle V_x\rangle/dF_D = 0.0$
even though the drive is increasing,
which indicates that the velocity in the $x$ direction
has ceased to increase with increasing $F_D$.
The $d\langle V_y\rangle/dF_D$ curves in
Fig.~\ref{fig:9}(b) all begin at a zero value
and also show a single strong
dip for small $\alpha_m/\alpha_d$, while a weaker second dip feature begins
to emerge as $\alpha_m/\alpha_d$ increases. At high $\alpha_m/\alpha_d$,
this second peak reaches its maximum magnitude and then decreases in
magnitude, while the original peak becomes more prominent again and
drops to lower values of $F_D$.
In Fig.~\ref{fig:9}(c),
the skyrmion Hall angle $\langle \theta_{sk}\rangle$
shows a saturation for $F_D > 1.0$,
indicating that the strong drive dependence of the Hall angle
is the most pronounced when the maximum pinning force is greater
than the driving force.

\begin{figure}
\includegraphics[width=\columnwidth]{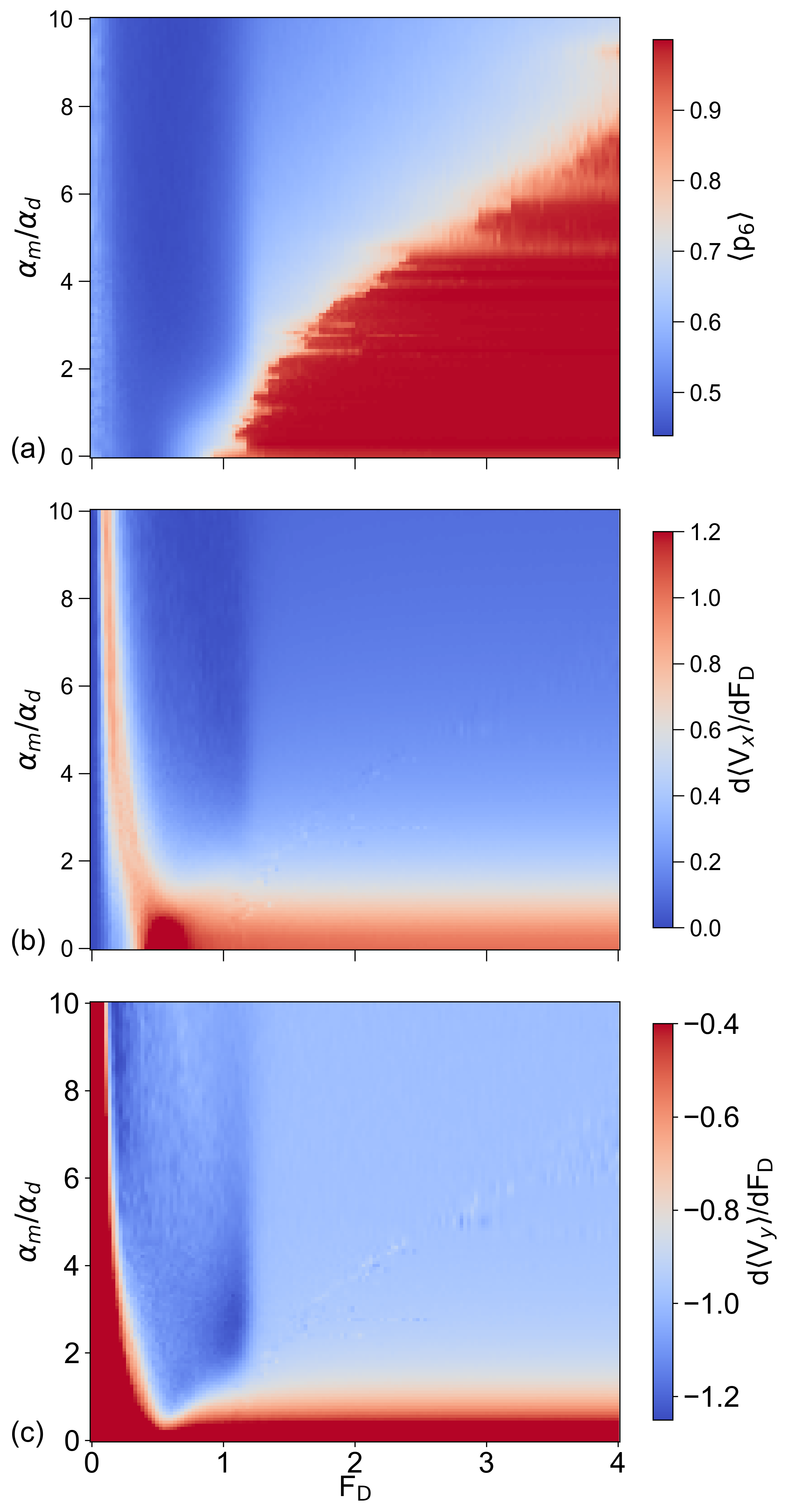}
\caption{Height fields as a function of $\alpha_m/\alpha_d$ vs $F_D$:
(a) $\langle p_6\rangle$, where red indicates the dynamic reordering.
(b) $d\langle V_x\rangle/dF_D$ showing the shifting of the main peak.
(c) $d\langle V_y\rangle/dF_D$, showing the evolution of the two minima.
}
\label{fig:10}
\end{figure}

In Fig.~\ref{fig:10}(a), we plot a height field of
the topological order $\langle p_6\rangle$ as a function of
$\alpha_m/\alpha_d$ versus $F_D$.
The red region indicates the formation of a dynamically ordered state in a
transition that shifts to higher
$F_D$ with increasing $\alpha_m/\alpha_d$.
The dark blue region is where the system is
the most strongly disordered, which occurs for $F_D = F_D/F_p < 1.0$
in the regime where the pinning is dominant.
There is also an intermediate or partially ordered regime
with $0.6 < \langle p_6\rangle < 0.8$ 
for $\alpha_m/\alpha_d > 2.0$.
This regime emerges when the Magnus force
becomes strong enough to induce
local rotations and topological defects even when all of the skyrmions
are flowing.
Figure~\ref{fig:10}(b) shows the
corresponding height field of $d\langle V_x\rangle/dF_D$
as a function of $\alpha_m/\alpha_d$ versus $F_D$.
There is a sharp peak above depinning
when $\alpha_m/\alpha_d < 1.0$, similar to that
observed in superconducting
vortex systems with $\alpha_m/\alpha_d = 0.0$ \cite{Reichhardt25}. 
For increasing $\alpha_m/\alpha_d$, the peak above depinning
shifts to lower drives, and when
$\alpha_m/\alpha_d > 2.0$,
there is a region where $d\langle V_x\rangle/dF_D = 0.0$ in which
the $x$ direction velocity does not increase with increasing $F_D$.
In Fig.~\ref{fig:10}(c), the corresponding
$d\langle V_y\rangle/dF_D$ height field shows that
the trends reverse, with small minima appearing
for low $\alpha_m/\alpha_d$ that become
stronger with increasing $\alpha_m/\alpha_d$.
Two minima are present when $\alpha_m/\alpha_d > 1.0$, but
at high $\alpha_m/\alpha_d$ the minimum near $F_D=1.0$ diminishes in
size and the minimum near the depinning transition becomes dominant.

\begin{figure}
\includegraphics[width=\columnwidth]{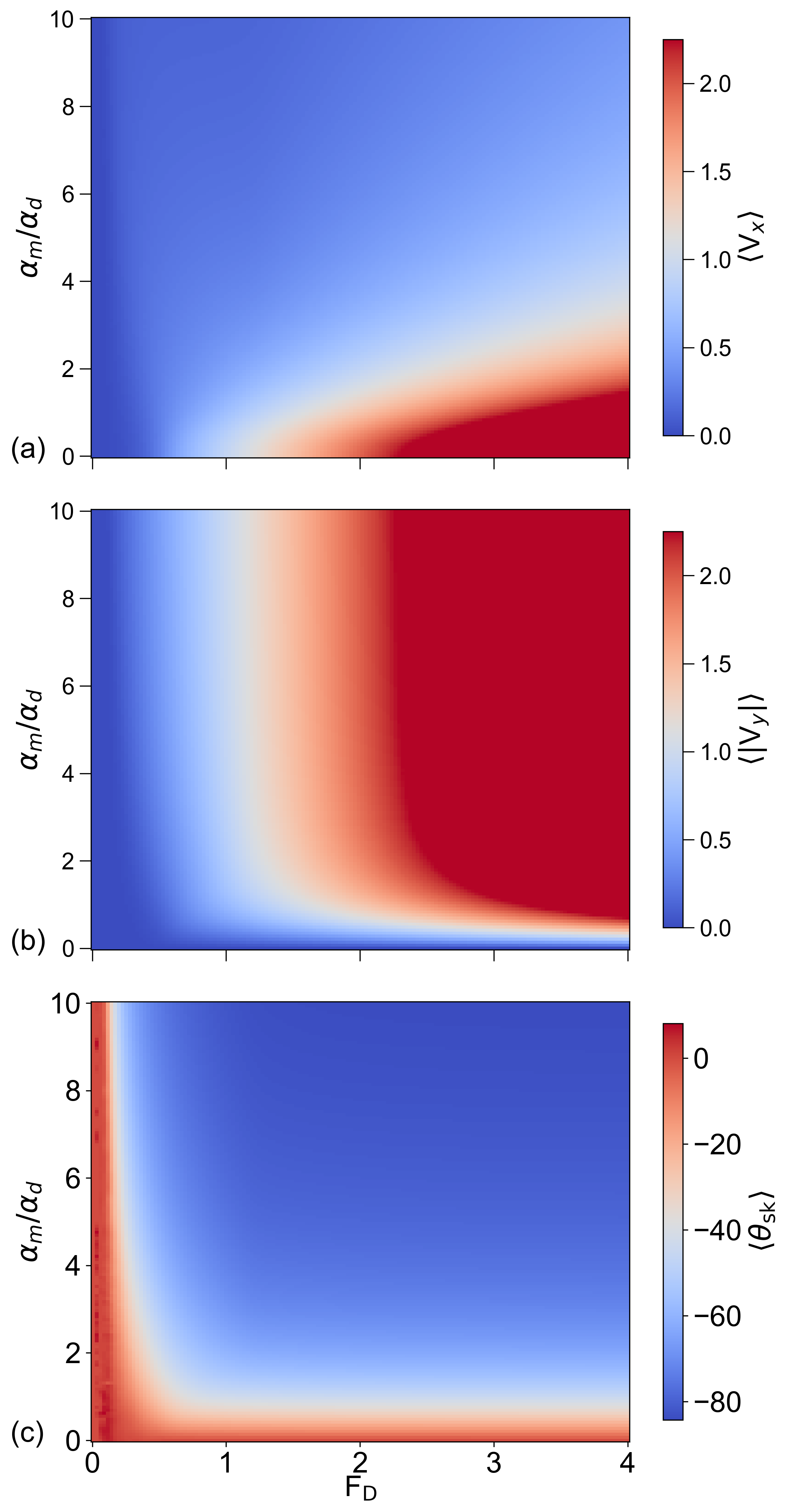}
\caption{Height fields as a function of $\alpha_m/\alpha_d$ vs $F_D$:
(a) $\langle V_x\rangle$. 
(b) $\langle|V_{y}|\rangle$.
(c) The skyrmion Hall angle $\langle \theta_{sk}\rangle$
showing lower values of $\langle \theta_{sk}\rangle$ near depinning.
}
\label{fig:11}
\end{figure}

In Fig.~\ref{fig:11}(a,b,c), we show the evolution of
$\langle V_x\rangle$, $\langle |V_y|\rangle$, and the skyrmion Hall angle
$\langle \theta_{sk}\rangle$, respectively, in the form of height fields
as a function of $\alpha_m/\alpha_d$ versus $F_D$.
The skyrmion Hall angle is small for $F_D < 0.5$
where the pinning is dominant,
while $\langle |V_y|\rangle$ increases
and $\langle V_x\rangle$ decreases
with increasing $F_D$.

\begin{figure}
\includegraphics[width=\columnwidth]{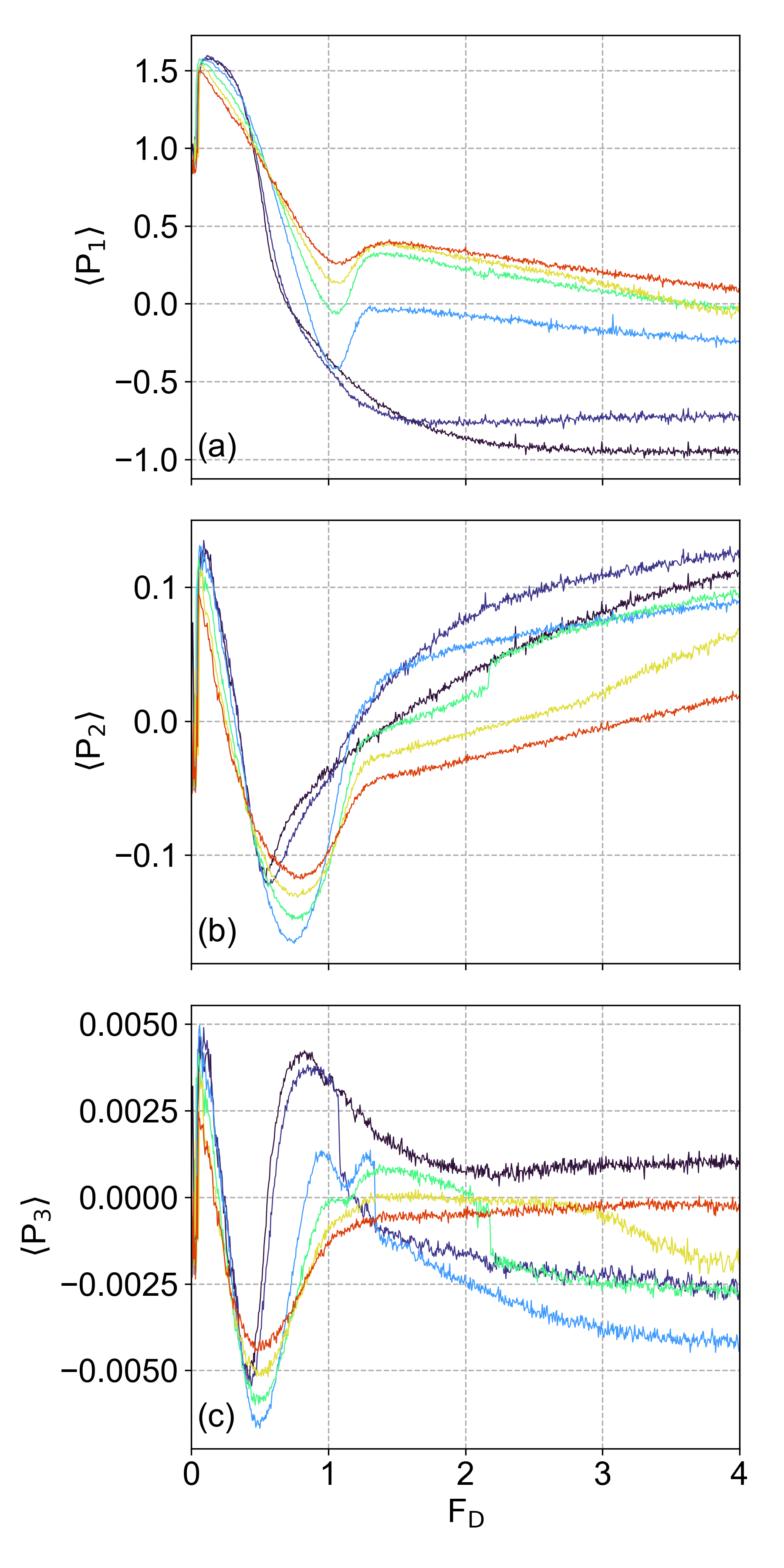}
\caption{Order parameters taken from the first three principal components
for systems with  
$\alpha_m/\alpha_d = 0.0$ (black), 0.5 (purple),
2.0 (blue), 4.0 (green), 6.0 (yellow), and 8.5 (red).
(a) $\langle P_{1}\rangle$ vs $F_{D}$.
(b) $\langle P_{2}\rangle$ vs $F_{D}.$
(c) $\langle P_{3}\rangle$ vs $F_{D}$. 
}
\label{fig:12}
\end{figure}

\subsection{PVB PCA Order Parameters}

We next consider order parameters derived from the PVB PCA,
where we specifically focus on averaged values of the first three
principal components, $\langle P_1\rangle$, $\langle P_2\rangle$,
and $\langle P_3\rangle$.
In Fig.~\ref{fig:12}(a,b,c),
we plot $\langle P_1\rangle$, $\langle P_2\rangle$,
and $\langle P_3\rangle$, respectively,
versus $F_D$ for
systems with $\alpha_m/\alpha_d = 0.0$, 0.5, 2.0, 4.0, 6.0, and 8.5.
For $\alpha_m/\alpha_d = 0.0$ and 0.5,
$\langle P_1\rangle$ drops from its initial peak at depinning,
crosses zero at about $F_D=0.7$ at the transition from
phase V to phase VI, and continues to decrease until
saturating at a drive near the crossover to the DR
phase, which is a
moving smectic for $\alpha_m/\alpha_d=0.0$ and a
dynamically reordered  lattice for $\alpha_m/\alpha_d=0.5$.
For the same systems,
$\langle P_2\rangle$ passes through a minimum at
the boundary between phases IV and V
and then crosses zero again
at the onset of the DR state.
$\langle P_3\rangle$ shows a peak at the depinning transition from phase I to phase II,
crosses zero at the crossover from phase II to phase III, and passes through
a dip at the boundary between phase III and phase IV.

For $\alpha_m/\alpha_d=2.0$ and $\alpha_m/\alpha_d = 4.0$ in Fig.~\ref{fig:12},
there is now a minimum in $\langle P_1\rangle$ near $F_D=1.0$
corresponding to a boundary between phase VI and a previously unobserved
moving liquid (ML) state, described in further detail below. For
$\alpha_m/\alpha_d=2.0$, $\langle P_1\rangle$ remains below zero at
higher drives, but for $\alpha_m/\alpha_d=4.0$, there is a new higher drive
zero crossing of $\langle P_1\rangle$ near $F_D=1.1$ that marks a
boundary between the ML state and a second previously unobserved
state, the clustered liquid (CL), where partial ordering has emerged but
the system has not dynamically reordered. Additional details about the CL
state appear below.
The minimum in $\langle P_2\rangle$ marking the IV-V boundary
deepens and
continues to shift to slightly
higher $F_D$ with increasing $\alpha_m/\alpha_d$.
At the same time, the zero crossing
of $\langle P_2\rangle$
that coincides with the appearance of the DR state, which had originally
dropped to lower $F_D$ between $\alpha_m/\alpha_d=0.0$ and
$\alpha_m/\alpha_d=0.5$, now moves back up to higher $F_D$ between
$\alpha_m/\alpha_d=2.0$ and $\alpha_m/\alpha_d=4.0$, marking non-monotonic
behavior of the DR boundary.
The $\langle P_3\rangle$ curve picks up a new zero crossing and a new peak
that coincide with the features in $\langle P_1\rangle$.

In the $\alpha_m/\alpha_d=6.0$ and 8.5 curves in Fig.~\ref{fig:12},
$\langle P_1\rangle$ retains a minimum near $F_D=1.0$ but no longer crosses
zero near the minimum. This corresponds to the disappearance
of phase VI, which is replaced by a direct boundary between phase V
and the ML state. The dip at the upper edge of phase V and the peak or
shoulder at the ML-MC boundary become less pronounced with increasing
$\alpha_m/\alpha_d$ but remain at nearly the same values of $F_D$.
The minimum in $\langle P_2\rangle$ at the IV-V boundary also remains at a
nearly constant value of $F_{D}$ for higher $\alpha_m/\alpha_d$, but the
upper zero crossing of $\langle P_2\rangle$ marking the CL-DR boundary
continues to shift rapidly to higher drives as $\alpha_m/\alpha_d$ increases.
The features of $\langle P_3\rangle$ show little dependence on
$\alpha_m/\alpha_d$ at the larger $\alpha_m/\alpha_d$ values.

\begin{figure}
\includegraphics[width=\columnwidth]{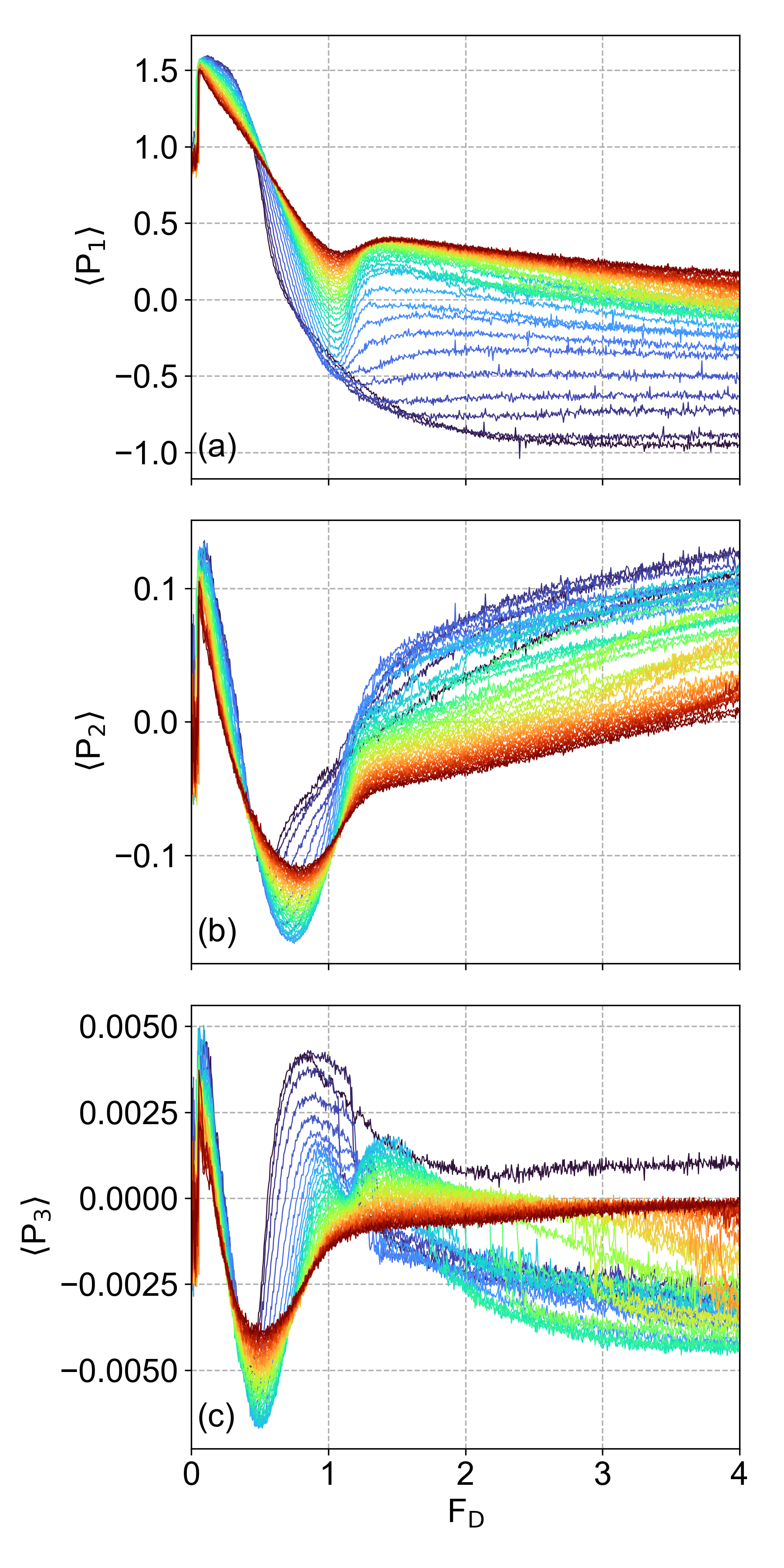}
\caption{Order parameters taken from the first three principal components
at varied $\alpha_m/\alpha_d$ from $\alpha_m/\alpha_d=0.0$ (black) to
$\alpha_m/\alpha_d=10.0$ (dark red) in intervals of 0.25.
(a) $\langle P_{1}\rangle$ vs $F_{D}$.
(b) $\langle P_{2}\rangle$ vs $F_{D}$.
(c) $\langle P_{3}\rangle$ vs $F_{D}$. 
}
\label{fig:13}
\end{figure}

\begin{figure}
  \includegraphics[width=\columnwidth]{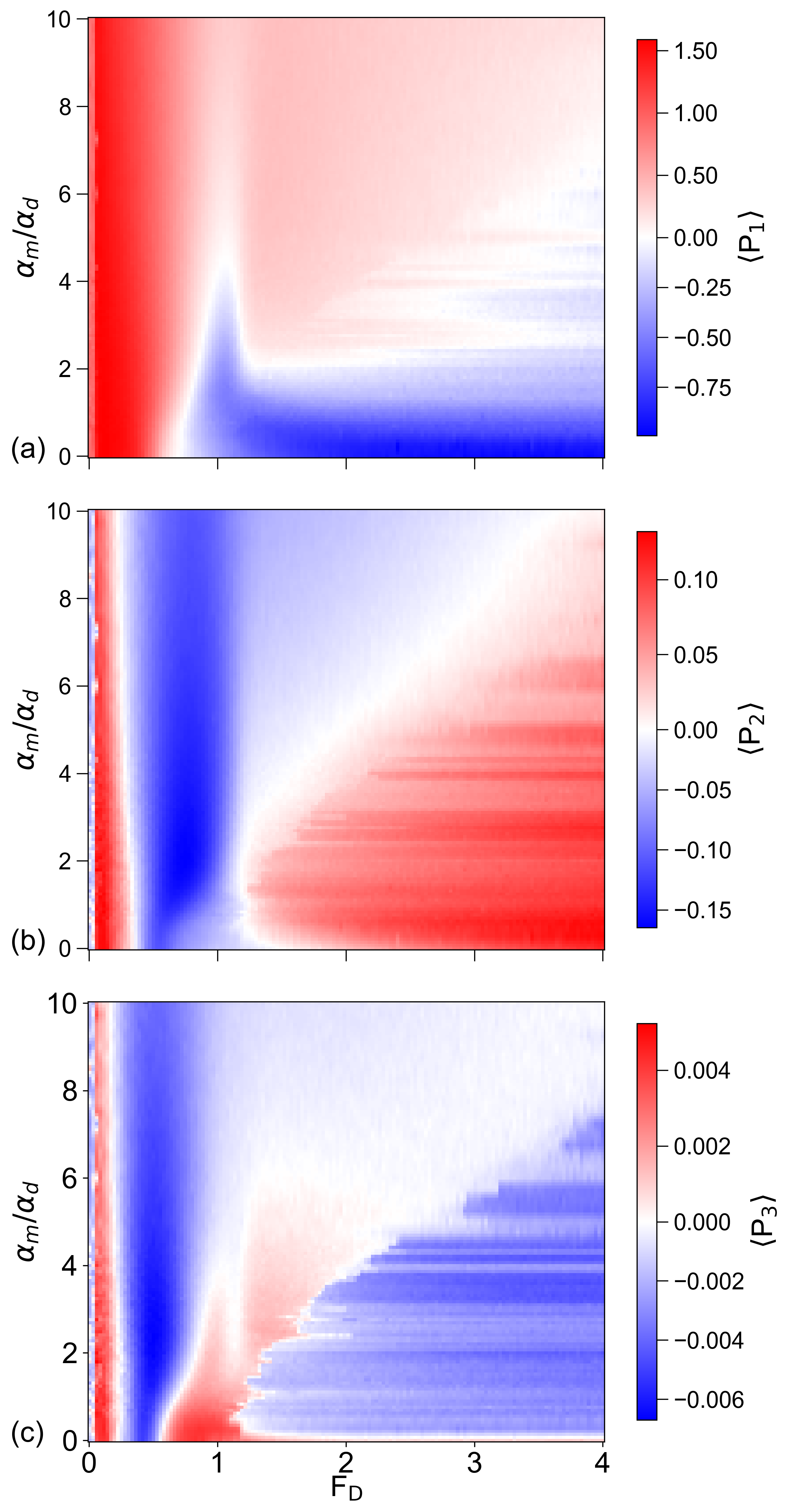}
  \caption{Height fields as a function of $\alpha_m/\alpha_d$ vs $F_D$:
(a) $\langle P_1\rangle$. (b) $\langle P_2\rangle$. (c) $\langle P_3\rangle$.     }
  \label{fig:NEW}
\end{figure}

To illustrate the full evolution of the PCA order parameters,
in Fig.~\ref{fig:13}(a,b,c), we plot
$\langle P_1\rangle$, $\langle P_2\rangle$, and
$\langle P_3\rangle$, respectively,
versus $F_D$ for $\alpha_m/\alpha_d$ ranging from 0 to 10.0 in
intervals of 0.25.
This information is also presented as height fields as a function of
$\alpha_m/\alpha_d$ versus $F_D$ in
Fig.~\ref{fig:NEW}.
With increasing $\alpha_m/\alpha_d$, $\langle P_1\rangle$
first develops a local minimum near
$F_D=1.0$ and then develops a local maximum at slightly higher drives, with both
features becoming less prominent at higher $\alpha_m/\alpha_d$, and ceasing to
cross zero for $\alpha_m/\alpha_d>4.25$.
The minimum in $\langle P_2\rangle$ shifts to higher $F_D$ with
increasing $\alpha_m/\alpha_d$ while first increasing and then decreasing
in depth. At the same time, the higher drive zero crossing continues to
climb to higher drives as $\alpha_m/\alpha_d$ becomes large.
The $\langle P_3\rangle$ curves retain the step-like jump feature found
in $\langle p_6\rangle$ at the transition to a more ordered state,
but also have a number of broader signatures that generally overlap with
signatures in $\langle P_2\rangle$ and $\langle P_1\rangle$.
The peak at depinning, which is followed by a zero crossing and then a
dip, are all fairly insensitive to $\alpha_m/\alpha_d$, with the dip becoming
more shallow as $\alpha_m/\alpha_d$ increases.

\begin{figure}
\includegraphics[width=\columnwidth]{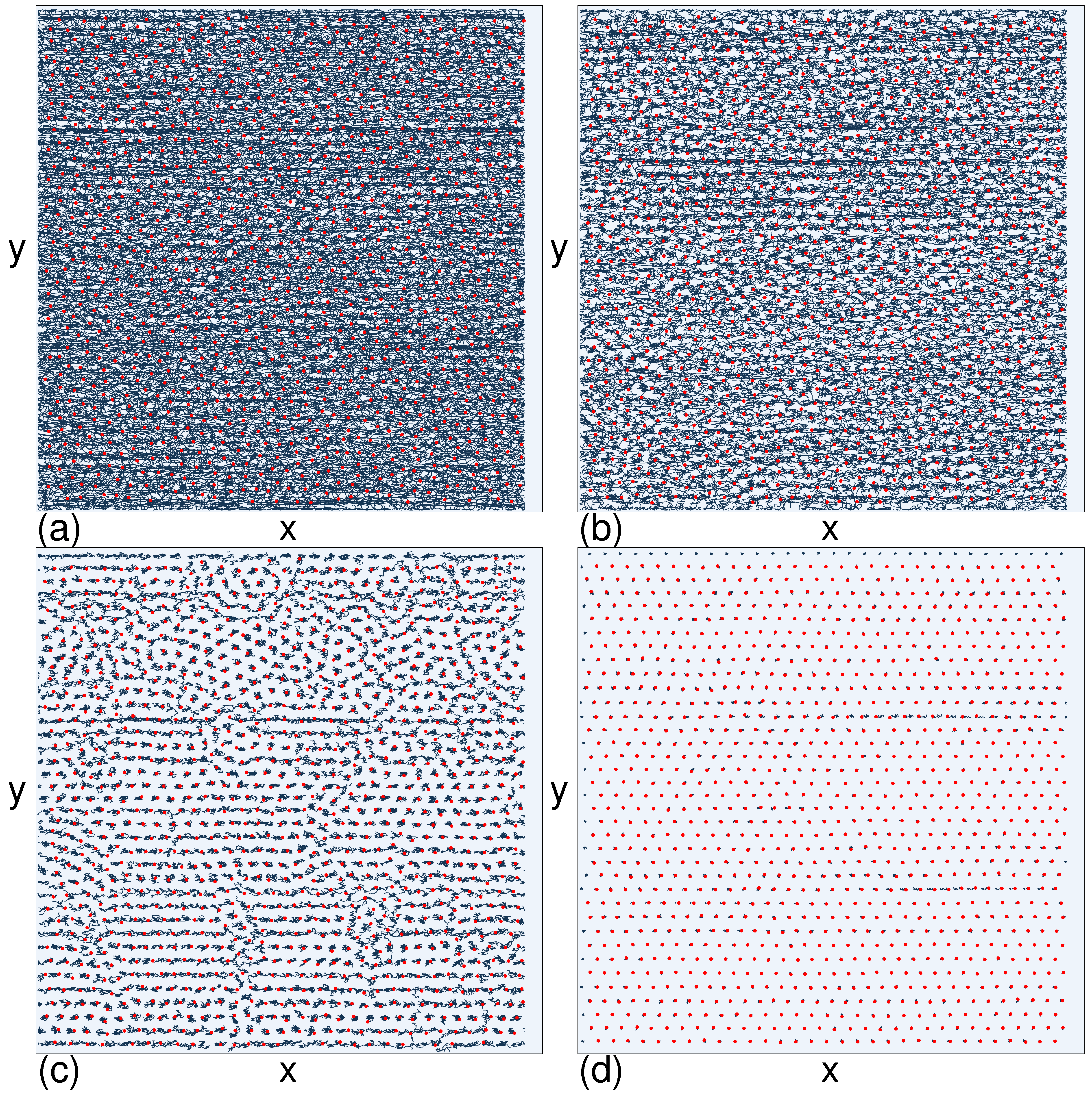}
\caption{Particle positions (dots) and trajectories (lines) in the moving
reference frame for a system in the superconducting vortex limit at
$\alpha_m/\alpha_d=0.0$.
As in the rest of this work, the trajectories are imaged over a time
equal to that required for an individual particle moving through a pin-free
sample to travel a distance of 50$\lambda$ when subjected to the
driving force $F_D$.
(a) The heavily braided channel phase IV at $F_D=0.48$, where the ordering
is still fluid-like but the displacements of the particles are anisotropic
and more prominent along the driving direction.
(b) The inhomogeneous ergodic plastic flow phase V at $F_D=0.61$. The system
is still fluid but is developing small locally ordered patches.
(c) The emerging ordered flow phase VI at $F_D=0.85$, where there is more
noticeable
ordering and local regions exhibit largely 1D diffusion.
(d) The moving smectic state or DR regime at $F_D=2.0$, where the ordering is
largely triangular but some aligned dislocations remain and produce strictly
1D motion along the rows.
}
\label{fig:14}
\end{figure}

In order to understand what is happening in the ML and CL states newly
detected by PVB PCA, we first made trajectory plots similar to those
shown in Figs.~\ref{fig:4}, \ref{fig:5}, and \ref{fig:6}. We found, however,
that unless the flow at higher drives is relatively ordered, as in the DR
state, the trajectories present an indistinguishable blurred mass.
Thus, we performed imaging in a moving reference frame in order to detect
the deviations each particle makes around its original position as a function
of time. This measure works badly when $F_D \leq 1.0$ because the pinning
remains dominant and, in the moving reference frame, produces a series of
streaks across the image. As the skyrmion structure increasingly begins to
decouple from the pinning
with increasing $F_D$, the moving reference
frame images become more informative.
Due to the rotation of the skyrmion direction of travel
by the Magnus force, we cannot simply subtract off the average velocity
in the $x$ direction with increasing $F_D$. Instead, we find the average
motion of the center of mass across all frames for a given current, and
subtract this average motion from all particles.
In Fig.~\ref{fig:14}(a-d), we show the moving frame images for
$\alpha_m/\alpha_d = 0.0$ (the superconducting vortex limit)
in phase IV at $F_D=0.48$ in Fig.~\ref{fig:14}(a),
phase V at $F_D=0.61$ in Fig.~\ref{fig:14}(b),
phase VI at $F_D=0.85$ in Fig.~\ref{fig:14}(c),
and the moving smectic or DR state at $F_D=2.0$ in
Fig.~\ref{fig:14}(d).
For phase IV, the ordering is fluid-like
but the diffusion is anisotropic and motion along the driving direction
is favored. Some pinned particles are
still present in phase IV and they produce streaks in the image.
Phase V has a similar overall liquid-like ordering,
but small patches of triangular ordering are beginning to emerge. Here
the streaks are produced by the slowing of particles passing through pinning
sites, rather than permanently pinned particles.
In phase VI the particles are much more ordered
and exhibit local regions of diffusion that has a strongly 1D character.
The DR or moving smectic state has well developed crystalline ordering
and shows localized areas
of 1D motion produced by the aligned dislocations that are present
in the smectic structure.

\begin{figure}
\includegraphics[width=\columnwidth]{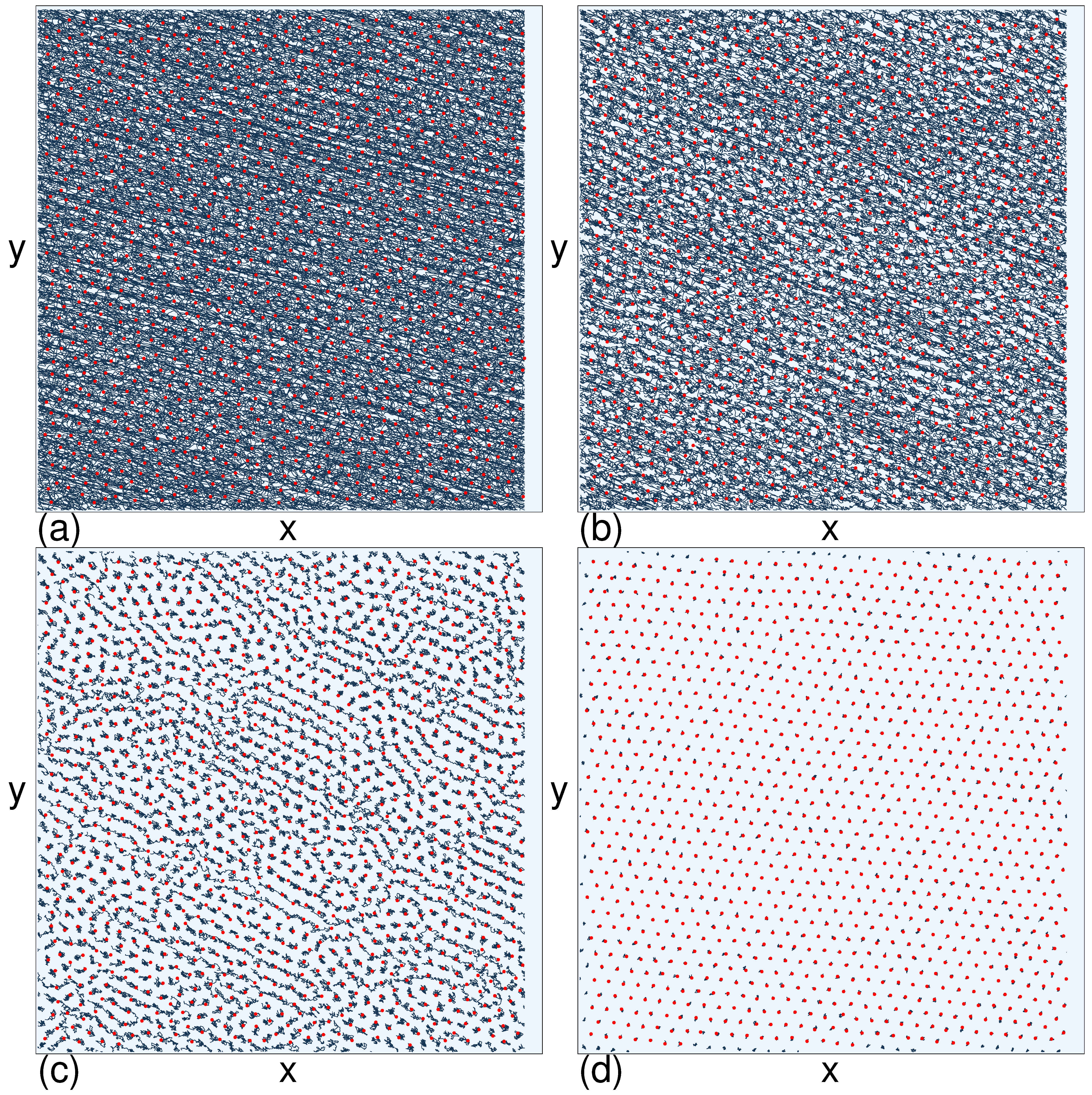}
\caption{Particle positions (dots) and trajectories (lines)
in the moving reference frame for a system with  
$\alpha_m/\alpha_d = 0.5$.
(a) The heavily braided channel phase IV at $F_D=0.48$, with fluid-like
structure.
(b) The inhomogeneous ergodic plastic flow phase V at
$F_D=0.63$, where the motion resembles an anisotropic liquid.
(c) The emerging ordered flow phase VI at $F_D=0.97$,
where greater order is present. (d) The moving crystal or DR phase at $F_D=2.0$.
}
\label{fig:15}
\end{figure}

\begin{figure}
\includegraphics[width=\columnwidth]{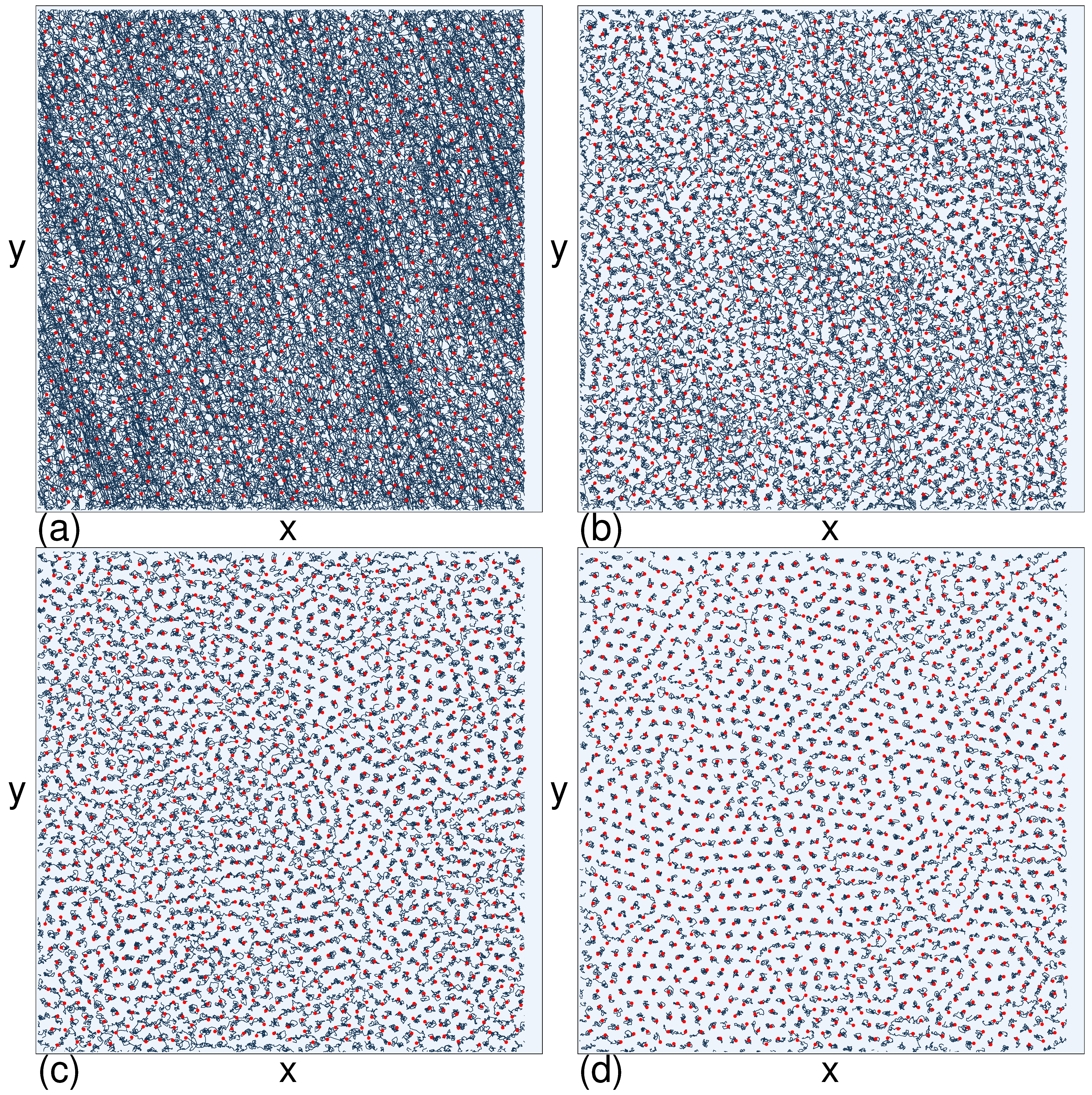}
\caption{Particle positions (dots) and trajectories (lines)
in the moving reference frame for a system with  
$\alpha_m/\alpha_d = 4.0$.
(a) The last vestiges of the emerging ordered flow
phase VI at $F_D=1.02$, where the motion has become relatively more
fluid.
(b) The moving liquid or ML phase at $F_D=1.22$, where the diffusion
is much more isotropic and the system in the moving frame resembles a
fluid.
(c) The clustered liquid or CL phase at $F_D=1.48$, where patches of
local triangular ordering are emerging and the diffusion resembles what would
be expected in an isotropic crystal above melting.
(d) Just into the DR state at $F_D=1.9$, where the diffusion is
becoming strongly localized and the crystalline ordering is improving.
}
\label{fig:16}
\end{figure}

Figure~\ref{fig:15} shows the moving reference frame images
at $\alpha_m/\alpha_d=0.5$ for phase IV at $F_D=0.48$ in Fig.~\ref{fig:15}(a),
phase V at $F_D=0.63$ in Fig.~\ref{fig:15}(b),
phase VI at $F_D=0.97$ in Fig.~\ref{fig:15}(c),
and the moving crystal or DR phase at $F_D=2.0$ in
Fig.~\ref{fig:15}(d).
In the presence of a finite Magnus term,
phases IV and V show strong diffusion, while
phase VI has more string-like diffusion, and
the DR phase is now a fully ordered crystal.
In Fig.~\ref{fig:16} we show the $\alpha_m/\alpha_d=4.0$ system in the
moving reference frame in
phase VI at $F_D=1.02$ in Fig.~\ref{fig:16}(a),
the moving liquid or ML phase at $F_D=1.22$ in Fig.~\ref{fig:16}(b),
the cluster liquid or CL phase at $F_D=1.48$ in Fig.~\ref{fig:16}(c),
and the DR phase at $F_D=1.9$ in Fig.~\ref{fig:16}(d).
This system is at the very edge of exhibiting phase IV; there is
still anisotropic transport present but the entire state is becoming
increasingly fluid-like.
In the ML phase the diffusion is now nearly isotropic, and the sequence
ML-CL-DR resembles the freezing of an isotropic crystal. The distinction
that PVB PCA draws between these three states is visible in the figure.
The ML phase has fluid-like order and fluid-like isotropic diffusion.
The CL phase has patches of local order and the diffusion has become
heterogeneous, with percolating strings of diffusion occurring in all
directions around small patches of more stationary particles.
Finally in the DR phase,
where some disorder persists due to the rotation of the particles
caused by the Magnus force, the diffusing strings have become isolated and
filamentary, and gradually pinch off as $F_D$ increases further.
Thus PVB PCA has successfully detected morphological changes in the nature
of the moving reference frame diffusion.

\begin{figure}
\includegraphics[width=\columnwidth]{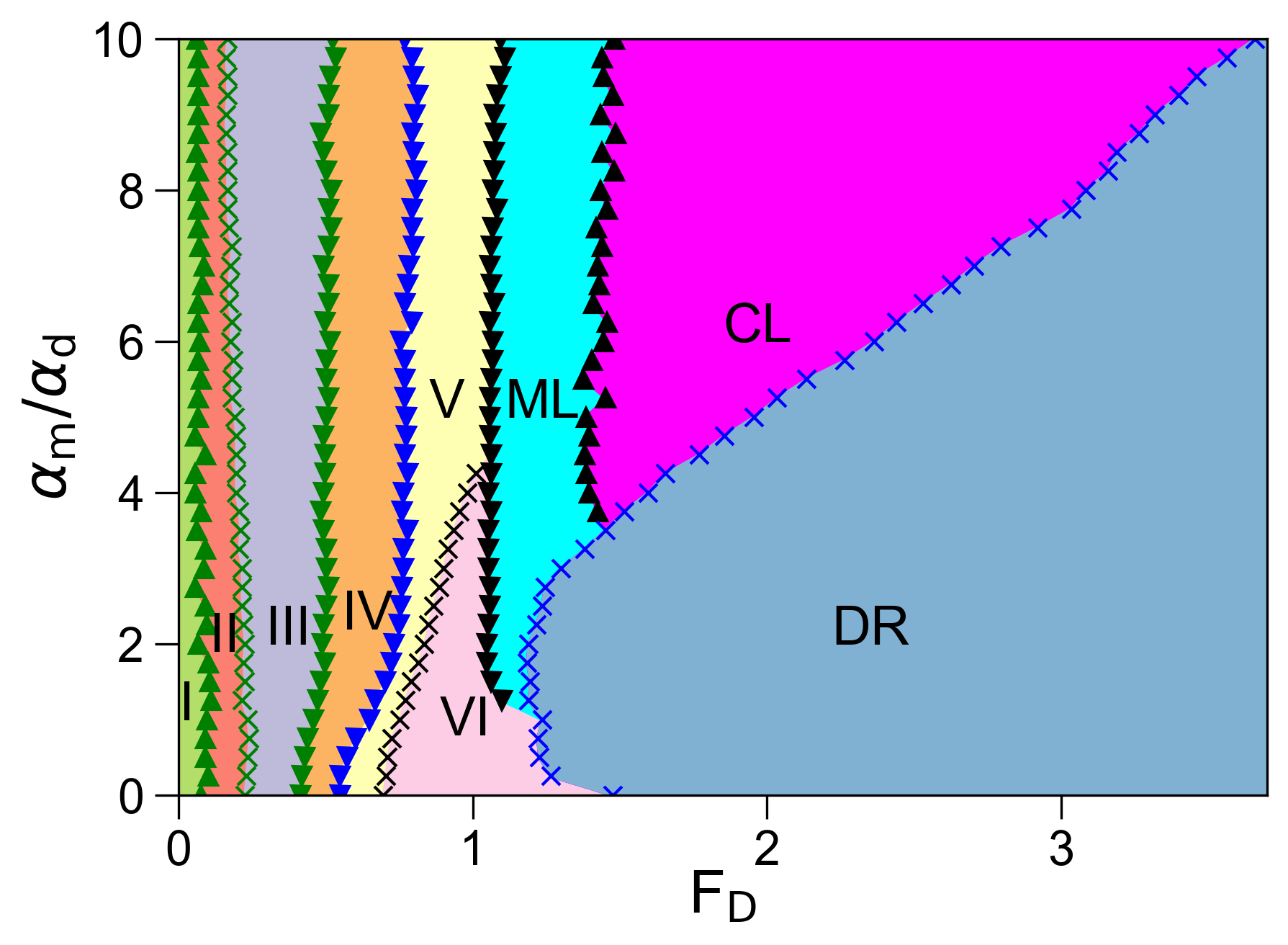}
\caption{Dynamic phase diagram as a function of $\alpha_m/\alpha_d$ vs $F_D$
constructed using the features identified by PVB PCA.
Green: Pinned (phase I).
Red: Isolated channel flow (phase II).
Purple: Lightly braided channel flow (phase III).
Orange: Heavily braided channel flow (phase IV) with a fraction of
permanently pinned particles.
Yellow: Inhomogeneous ergodic plastic flow (phase V) with no permanently
pinned particles.
Pink: Emerging ordered flow (phase VI).
Blue: Dynamically reordered flow (phase DR),
which is a moving crystal for finite Magnus force and a
moving smectic for zero Magnus force.
The phases newly detected by PVB PCA are:
Cyan: Moving liquid flow (phase ML) where all the particles are moving
and there is strong rotation due to the Magnus term.
Magenta: Clustered liquid flow (phase CL), a
partially ordered phase where the system tries to form a triangular lattice
but contains numerous
topological defects.
The phase boundaries are obtained from the following measures:
I-II (green up triangles): peak of $\langle P_3\rangle$, which also
coincides with the lowest minimum of $\langle P_1\rangle$.
II-III (green crosses): lowest zero crossing of $\langle P_3\rangle$.
III-IV (green down triangles): minimum of $\langle P_3\rangle$.
IV-V (blue down triangles): minimum of $\langle P_2\rangle$.
V-VI (black crosses): zero crossing of $\langle P_1\rangle$.
VI-ML (black down triangles): minimum of $\langle P_1\rangle$.
ML-CL (black up triangles): upper maximum of $\langle P_1\rangle$.
Boundary of DR phase (blue crosses):
upper zero crossing of $\langle P_2\rangle$.
Note that all order parameters are only defined to within a sign.
}
\label{fig:17}
\end{figure}

From the different features in $\langle P_1\rangle$, $\langle P_2\rangle$,
and $\langle P_3\rangle$,
we can construct a dynamical phase diagram as shown in
Fig.~\ref{fig:17} as a function of $\alpha_m/\alpha_d$ versus
$F_{D}$. Here, phase I is the pinned phase,
phase II is the isolated 1D static channel phase,
and phase III is the lightly braided quasi-2D channel phase.
The skyrmion Hall angle is close to zero in phase II and low in phase III.
Phase IV is a heavily braided channel flow state in which
a portion of the particles remain permanently pinned,
so that the flow is still non-ergodic.
Phase V is the disordered, inhomogeneous ergodic plastic flow phase,
where some particles are temporarily pinned,
but all of the particles take part in the motion.
Phase VI is an emerging ordered flow state in which all of the particles
are moving at close to the same speed but the system remains disordered.
Phase ML is a disordered moving liquid phase
in which all of the particles are moving,
the diffusion in the moving lattice frame is
isotropic, and there is strong rotation of the particle
trajectories due to the Magnus phase.
Phase CL is the partially ordered cluster liquid
phase where the system tries to form a triangular lattice,
but the local velocity differences produced by the pinning
and the strong Magnus force create local rotations
which induce the formation of some topological defects.
DR is the dynamically ordered phase,
which is a crystal for finite Magnus force and a smectic for zero Magnus force.

\section{Discussion}

We have focused on a single value of the pinning force $F_p$
which is large enough that the depinning process is plastic.
For weaker pinning, the system should enter an elastic depinning
regime in which each particle
keeps the same neighbors during depinning and sliding \cite{Reichhardt22a}.
In the elastic depinning regime, it
has been shown that the skyrmion Hall angle still
has a drive dependence, and
that this drive dependence becomes more pronounced where
thermal effects are introduced.
For our stronger pinning, if thermal effects are included
in the plastic depinning regime,
it is likely that phase I would be replaced by phase II, and
that phases III, IV, and V would be shifted to lower values of $F_D$.

A broader question is whether the boundaries between the
phases detected by PVB PCA are nonequilibrium
phase transitions or crossovers.
Previous work provided evidence that the plastic depinning
transition is a nonequilibrium phase transition with
critical properties \cite{Reichhardt17}, so it could be possible
that at least some of the other phase boundaries are also marked by
nonequilibrium phase transitions.
In particular, the crossover from non-ergodic to ergodic plastic flow
at the IV-V boundary
could fall in the class of absorbing phase transitions
\cite{Hinrichsen00,Corte08}, and the
dynamical reordering transitions could represent
another type of nonequilibrium phase transition.
It may be that the II-III or III-IV transitions
are examples of different kinds of directed percolation
with different quantities percolating.
The percolation of diffusion in the moving reference frame could show
similar transitions at the V-ML, ML-CL, and CL-DR boundaries.
Although the PVB PCA can find evidence for the
presence of different phases, it does not determine whether the
boundaries it marks are nonequilibrium phase transitions.
Future directions include determining whether the PVB PCA-derived order
parameters themselves show scaling,
and if they do, whether the exponents could be related to established
universality classes.

\section{Summary}
We have investigated the dynamics of driven skyrmions
using a modified Thiele equation approach and a combination of
standard measures such as features in the velocity-force curves,
differential conductivity, and fraction of topological defects in combination
with a position-and-velocity based principal component analysis (PVB PCA).
The PVB PCA combines information about both the
spatial and velocity properties
of the moving particles.
The standard measures show evidence for plastic depinning and
dynamic reordering, but from these measures alone,
it is difficult to determine whether
additional dynamic phases occur in the plastic flow regimes.
Principal component analysis was previously used to distinguish
different types of plastic flow regimes for driven overdamped systems
moving over random disorder.
Here, the strong Magnus component of the
skyrmion dynamics produces previously unobserved higher velocity
flow phases.
For the skyrmions, we find that just above depinning,
the flow follows static one-dimensional channels
and that the skyrmion Hall angle is zero.
At higher drives, the channels become lightly braided around
permanently pinned particles,
and the skyrmion Hall angle becomes finite but remains small.
As the drive increases further,
the channel flow
becomes more two-dimensional and strongly fluctuating, but there are
still some permanently pinned particles, so the system is in a
non-ergodic regime and the flow is described as heavily braided
channel motion.
Once the driving force exceeds the pinning force, we find
a regime in which all of the particles take part
in the motion over time, but some particles can be temporarily immobile,
giving an inhomogeneous ergodic plastic flow state.
At higher drives, in addition to the emerging ordered flow state found
for overdamped particles, we observe two previously unreported
plastic flow states: the moving liquid and the cluster liquid, in which
the system when viewed in the moving reference frame takes on the appearance
of a melted crystal that is being cooled.
At the highest drives, the system can dynamically order
into a moving crystal.
Using the first three principal components as order parameters,
we map out the evolution of these different phases as
a function of the ratio of the Magnus to the
damping force as well as the external drive.
For increased Magnus force, there is an extended region of
partially ordered flow phases that arise
due to local rotations caused by the pinning.
Our results indicate that PVB PCA,
which takes into account both positions and velocities,
can be used to distinguish different dynamical phases
more accurately than standard transport
measurements or changes in the defect density.
The PVB PCA approach can be used as a general tool
that is applicable to other
types of driven systems that could exhibit different dynamical phases.

\acknowledgments
We gratefully acknowledge the support of the U.S. Department of
Energy through the LANL/LDRD program for this work.
  This work was supported by the US Department of Energy through
  the Los Alamos National Laboratory.  Los Alamos National Laboratory is
  operated by Triad National Security, LLC, for the National Nuclear Security
  Administration of the U. S. Department of Energy (Contract No. 892333218NCA000001).
  
\bibliography{mybib}

\end{document}